\documentclass[particles,article,submit,pdftex,oneauthor]{Definitions/mdpi} 
\firstpage{1} 
\pubvolume{1}
\issuenum{1}
\articlenumber{0}
\pubyear{2025}
\copyrightyear{2025}
\datereceived{ } 
\daterevised{ } 
\dateaccepted{ } 
\datepublished{ } 
\hreflink{https://doi.org/} 
\pdfoutput=1 

\usepackage{dcolumn}
\usepackage{bm}
\usepackage[dvipsnames]{xcolor}

\newcommand{\rmd}{\mathrm{d}}
\def\araa{Annual Reviews of Astronomy and Astrophysics}
\def\ssr{Space Science Reviews}
\def\physrep{Physics Reports}
\def\nphysa{Nuclear Physics A}
\def\prd{Physical Review D}
\def\prc{Physical Review C}
\def\prl{Physical Review Letters}
\def\mnras{Monthly Notices of Royal Astronomical Society}

\def\apj{The Astrophysical Journal}

\Title{Neutrino pair bremsstrahlung due to electromagnetic collisions in neutron star cores revisited}

\TitleCitation{Neutrino pair bremsstrahlung due to electromagnetic collisions in neutron star cores revisited}

\Author{Peter S. Shternin$^{1}$*\orcidA{}}

\AuthorNames{Peter S. Shternin}

\AuthorCitation{Shternin, P.S.}

\address[1]{
Ioffe Institute, Politekhnicheskaya 26, St. Petersburg, 194021 Russia; pshternin@gmail.com
}

\corres{Correspondence: pshternin@gmail.com; }

\abstract{We reconsider the problem of neutrino pair bremsstrahlung emission originating from the electromagnetic collisions of charged particles in nucleonic ($npe\mu$) neutron star cores. Two limiting cases are considered: (i) protons are in the normal state and 
(ii) protons are in the superconducting state. In both cases, the dominant contribution to the  bremsstrahlung emissivity $Q^{\mathrm{em}}_{\mathrm{Br}}$ comes from the transverse part of in-medium electromagnetic interactions. For non-superconducting matter, we obtain an unusual $Q^{\mathrm{em}}_{\mathrm{Br}}\propto T^{23/3}$ temperature dependence due to the dynamical character of  plasma screening in the transverse channel, but considerably smaller values of $Q^{\mathrm{em}}_{\mathrm{Br}}$ than in previous studies,  rendering the considered process unimportant in practice. In contrast, for superconducting and superfluid matter,
the neutrino emission processes involving nucleons are suppressed and $Q^{\mathrm{em}}_{\mathrm{Br}}$ due to lepton collisions provides the residual contribution to the neutrino emissivity of neutron star core matter. In the superconducting case, the plasma screening becomes static and the standard $Q^{\mathrm{em}}_{\mathrm{Br}}\propto T^{8}$ temperature scaling is restored. Simple analytical expressions for  $Q^{\mathrm{em}}_{\mathrm{Br}}$ in both limiting cases are provided. 
}


\begin{document}


\section{Introduction}

Neutrino emission from neutron star (NS) interiors plays a pivotal role in their thermal evolution, at least during the first Myr of their life \cite{YakovlevPethick2004ARA&A, Potekhin2015SSRv}. There is a wealth of possible neutrino generation processes that can operate in NSs, depending on the composition and physical conditions \cite{YakovlevKaminker,SchmittShternin2018}. The dominant processes are usually the baryon Urca processes --- a pair of baryon beta-decay and inverse reactions --- operating as is (direct Urca) or in the presence of auxiliary particles (modified Urca). Additionally, neutrino pair bremsstrahlung in baryon collisions mediated by strong interactions may be significant. However, there are other, subdominant processes which may be important provided the favorable conditions are fulfilled.   

In the present study, we focus on one example of such reactions, namely neutrino pair bremsstrahlung emission from collisions of charged particles, mainly leptons, mediated by electromagnetic interactions. These reactions can be important in the regions of the star where reactions involving baryons are strongly suppressed by pairing \cite{YakovlevKaminker}.  

Electron-electron (\textit{ee}) and electron-proton (\textit{ep}) collisions in nucleonic NS cores in normal (that is, non-superconducting) matter and in the presence of proton pairing were considered by \citet{Kaminker1999}. Dense medium polarization effects were included via the static Thomas-Fermi screening of the electromagnetic interaction. It was later noted by \citet{Jaikumar2005PhRvD} that this approach is not  quite accurate in the relativistic plasma where the transverse part of the interaction is important which in normal matter is unscreened in the static limit. For low-temperature degenerate matter, this leads to a strong enhancement of the collision frequencies mediated by electromagnetic interactions and their 
unusual temperature dependence
\cite{Baymetal1990,HeiselbergPethick1993PhRvD}. 
The authors of Ref.~\cite{Jaikumar2005PhRvD} indeed found that the emissivity of \textit{ee} bremsstrahlung can be several orders of magnitude larger than the results of Ref.~\cite{Kaminker1999}. 
In the context of transport properties of QED and QCD matter, this effect was first pointed out and studied in \cite{Baymetal1990,Heiselberg1992,HeiselbergPethick1993PhRvD} and elaborated for  NS core physical conditions in a series of subsequent works; see Refs.~\cite{SchmittShternin2018,Shternin2022} for a review. The strong enhancement of \textit{ee} bremsstrahlung obtained by \citet{Jaikumar2005PhRvD} made the emissivity of this process comparable to nucleon bremsstrahlung rates mediated by strong  interactions in the normal (i.e., unpaired) matter. In the regions of the core where  strong neutron pairing suppresses the neutron contribution (that is, Urca reactions, bremsstrahlung reactions involving neutrons, and neutrino emission due to neutron Cooper pairing),  \textit{ee} bremsstrahlung turned out to be the dominant neutrino emission process over the concurrent \textit{pp} bremsstrahlung mediated by strong interactions.
The authors of Ref.~\cite{Jaikumar2005PhRvD} also briefly discussed the case of the simultaneous occurrence of strong proton and neutron pairing. They noted that in superconducting matter, the character of transverse screening changes from dynamical to static as the transverse photons acquire the Meissner mass. This leads to a strong suppression of the \textit{ee} bremsstrahlung emissivity, lowering it back to the orders of magnitude obtained in Ref.~\cite{Kaminker1999}, with the Meissner mass replacing the  Thomas-Fermi screening momentum in the transverse part of the interaction.

Unfortunately, it turns out that the expression for the electron propagator, used in Ref.~\cite{Jaikumar2005PhRvD} for calculating the matrix element of the process in question, is not valid in the relativistic regime.
We found that if  the appropriate form of the electron propagator is used,  
a much weaker enhancement of the \textit{ee} bremsstrahlung emissivity is obtained.

Therefore, the aim of this report is to 
provide corrected expressions for neutrino emissivity via electromagnetic bremsstrahlung in NS cores.
We focus on the nucleonic npe$\mu$ composition of the core, although the generalization to other charged particle species is straightforward.  For completeness, we do not rely on either ultrarelativistic or non-relativistic limits (as in Refs.~\cite{Kaminker1999,Jaikumar2005PhRvD}). This   allows one to use the obtained expressions to describe processes with the participation of mildly-relativistic muons or non-relativistic protons.
We also consider the case of superconducting matter at low temperatures ($T\lesssim 0.3 T_{Cp}$ with $T_{Cp}$ being proton critical temperature) following Ref.~\cite{Shternin2018} where the related problem of transport coefficients in superconducting NS cores was studied. While the transverse channel of electromagnetic interaction is also found to provide the dominant contribution, we show that it is not enough to consider the momentum-independent Meissner screening. That is, in the inner core of a NS the momentum dependence of the transverse polarization function starts to be important. As a result, a dominant `transverse' contribution to bremsstrahlung emissivity does not depend on $T_{Cp}$ at low density and acquires a weak $T_{Cp}^{-1/3}$ dependence on proton critical temperature at higher densities.
We argue that only in the case of simultaneous presence of strong neutron and proton pairing may the bremsstrahlung processes governed by electromagnetic interaction be important for  the neutrino emission budget of the star.

The paper is organized as follows. In Sec.~\ref{sec:form} we describe the formalism used for calculation of the bremsstrahlung neutrino emissivity, closely following Ref.~\cite{Jaikumar2005PhRvD}. The cases of normal and superconducting protons are considered in Secs.~\ref{sec:screening_normal}  and \ref{sec:screening_sc}. We discuss the obtained results in Sec.~\ref{sec:discus}, and provide practical expressions for bremsstrahlung emissivity and conclude in Sec.~\ref{sec:concl}. 

Throughout the paper, unless otherwise stated, we use natural units with
$\hbar=c=k_B=1$. The metric tensor $g^{\mu\nu}$ has a signature $(+,-,-,-)$. The Levi-Civita symbol is $\epsilon^{\alpha\beta\gamma\delta}$, and a shorthand notation for contraction $\epsilon^{q\beta\gamma\delta}\equiv q_\alpha \epsilon^{\alpha\beta\gamma\delta}$, where $q$ is some four-vector, is employed.  Greek letter tensor indices run from $0$ to $3$, while Roman tensor indices denote spatial coordinates and run from $1$ to $3$. Dirac gamma matrices are $\gamma^\mu$ and $\gamma^5=i\gamma^0\gamma^1\gamma^2\gamma^3$.

\section{Formalism}\label{sec:form}
\subsection{General expressions}

Electrons and muons in NS cores form (almost) free relativistic Fermi gases. In contrast, neutrons and protons form non-ideal, strongly interacting plasma \cite{HPYa-book,OfengeimEncycl2026}. In the present study, we will assume that the relativistic Fermi-liquid description is appropriate \cite{BaymChin1976NuPhA,FrimanWeise2019PhRvC} and rely on the quasiparticle approximation in calculating the bremsstrahlung rates involving protons. We note at this point that non-Fermi liquid effects in neutrino emission processes can also be important, as a recent study shows \cite{Sedrakian2024PhRvL}.

Quite generally, the neutrino emissivity of dense matter can be obtained by means of the optical theorem through the calculation of the weak response of the medium \cite[e.g.,][]{Sedrakian2007PrPNP,SedrakianDieperink2000PhRvD,Leinson2001}. Alternatively, within the quasiparticle approximation, one can rely on Fermi's Golden Rule provided the in-medium modifications of quasiparticle interactions and propagators are properly accounted for. Here we will stick to the latter approach, allowing us to make a direct comparison and connection with the results of previous works \cite{Kaminker1999,Jaikumar2005PhRvD}.

Within the Fermi's Golden Rule approach, neutrino pair emissivity $Q^{\mathrm{em}}_{\mathrm{Br}}$ in the bremsstrahlung process $1+2\to 3+4+\nu+\overline{\nu}$ can be calculated as
\cite[e.g.,][]{Kaminker1999,Jaikumar2005PhRvD}
\begin{adjustwidth}{-\extralength}{0cm}
    \begin{equation}
        Q^{\mathrm{em}}_{\mathrm{Br}} = (2 \pi)^4\left[\prod\limits_{j = 1}^{4}\int \frac{\rmd^3 p_j}{(2 \pi)^3 2 E_j}\right] \int \frac{\rmd^3 q_1}{(2 \pi)^3 2 \omega_1} \frac{\rmd^3 q_2}{(2 \pi)^3 2 \omega_2} Q_0 
        \sum\limits_{\mathrm{spins}}\frac{|M|^2}{s} n_F(E_1)n_F(E_2) \widetilde{n}_F(E_3) \widetilde{n}_F(E_4)\  
        \delta^{(4)}(P_{\mathrm{fin}} - P_{\mathrm{in}}),\label{eq:Qbremms}
\end{equation}
\end{adjustwidth}
where the index $j=1\dots 4$ runs over the initial ($1,\, 2$) and final ($3,\ 4$) quasiparticle states, with 1 and 3 (2 and 4) corresponding to the same particle species. The first four 3D integrations are carried out over the quasiparticle three-momenta $\mathbf{p}_j$ with $P_j=(E_j,\,\mathbf{p}_j)$ being the corresponding four-momenta and $E_j$ the corresponding energies. The next two integrations are performed over the neutrino three-momenta $\mathbf{q}_{1,2}$, where neutrino four-momenta are $Q_{1,2}=(\omega_{1,2},\,\mathbf{q}_{1,2})$. The total neutrino pair four-momentum is denoted as $Q=Q_1+Q_2=(\omega_1+\omega_2,\mathbf{q}_1+\mathbf{q}_2)\equiv (Q_0,\mathbf{q})$. The neutrino pair energy $Q_0$ in Eq.~(\ref{eq:Qbremms}) ensures that the emissivity is calculated. To obtain the total emissivity of the bremsstrahlung process, Eq.~(\ref{eq:Qbremms}) needs to be summed over neutrino flavors which are emitted incoherently.

The functions $n_F(E_j)$ are the Fermi-Dirac distributions for quasiparticles,
\begin{equation}\label{eq:FD}
    n_F(E_j)=\frac{1}{\mathrm{e}^{(E_j-\mu_j)/T}+1},
\end{equation}
where $T$ is the temperature, $\mu_j$ is the chemical potential for particle species $j$, and in Eq.~(\ref{eq:Qbremms}) we have abbreviated $\widetilde{n}_F(E_j)\equiv 1-{n}_F(E_j)$.
In strongly degenerate matter of NS cores, chemical potentials are equal to the quasiparticle Fermi energies, $\mu_j=E_{\mathrm{F}j}=E_j(p_{\mathrm{F}j})$, where $p_{\mathrm{F}j}$ are the corresponding Fermi momenta, $p_{\mathrm{F}j}=(3\pi^2 n_j)^{1/3}$, and $n_j$ are the number densities of particle species $j$. For an almost free lepton gas, the dispersion relation for quasiparticles is simply $E_j(p_j)=\sqrt{p_j^2+m_j^2}$ where $m_j$ is the bare mass of particle species $j$ . For protons, we adopt the relativistic Fermi-liquid description \cite[e.g.,][]{FrimanWeise2019PhRvC} within the relativistic mean field \cite{Walecka1974AnPhy,DiracBrueckner1999IRvNP} scheme. That is, we neglect the quasiparticle self-energy dependence on momentum in a vicinity of the Fermi surface (at a given density $\rho$) and assume that the quasiparticle dispersion relation can be written as $E_j(p_j,\rho)=\widetilde{E}_j(p_j,\rho)+U_j(\rho)$, where $\widetilde{E}_j(p_j,\rho)=\sqrt{p_j^2+\widetilde{m}_j^2(\rho)}$ and $U_j(\rho)$ is a density-dependent mean-field potential \cite{FrimanWeise2019PhRvC}. Here, the quantity $\widetilde{m}_j$ is the so-called quasiparticle Dirac effective mass.
The quasiparticle Landau effective mass $m_j^*$ which defines the density of states  at the Fermi surface \cite{BaymChin1976NuPhA,BaymPethick91} is then just an `effective' Fermi energy $m_j^*=\widetilde{E}_{\mathrm{F}j}=
\sqrt{p_{\mathrm{F}j}^2+\widetilde{m}_j^2(\rho)}$. Indeed, defining the shifted chemical potential as $\widetilde{\mu}_j=\mu-U_j(\rho)$, the quasiparticle spectrum near the Fermi surface can be written as $E_j-\mu_j=\widetilde{E}_i-\widetilde{\mu}_j\approx v_{\mathrm{F}j}(p_j-p_{Fj})$, where the Fermi velocity is
\begin{equation}\label{eq:vF}
    v_{\mathrm{F}i}=\left(\frac{\partial E_i}{\partial p_{i}}\right)_{p_i=p_{\mathrm{F}i}}=\left(\frac{\partial \widetilde{E}_i}{\partial p_{i}}\right)_{p_i=p_{\mathrm{F}i}}=\frac{p_{\mathrm{F}i}}{\widetilde{E}_{\mathrm{F}i}}=\frac{p_{\mathrm{F}i}}{\sqrt{p_{\mathrm{F}i}^2+\widetilde{m}_{i}^2}}.
\end{equation}
In this approximation, we can proceed with Eq.~\eqref{eq:Qbremms} for collisions involving protons as well, 
substituting  $m_i\to \widetilde{m}_i$ and $E_i\to\widetilde{E}_i$ in the final expressions. We will achieve this goal by writing the final answer in terms of Fermi velocities and Fermi momenta.

The energy-momentum conserving delta-function $\delta^{(4)}(P_{\mathrm{fin}} - P_{\mathrm{in}})$ in Eq.~(\ref{eq:Qbremms}) contains the initial four-momentum $P_{\mathrm{in}}=P_1+P_2$ and the final four-momentum $P_{\mathrm{fin}}=P_3+P_4+Q$. It can be expressed as a product of the energy- and momentum-conserving delta functions, namely
\begin{equation}
    \delta^{(4)}(P_{\mathrm{fin}} - P_{\mathrm{in}})=
    \delta\left(E_3+E_4+Q_0-E_1-E_2\right)
\delta^{(3)}\left(\mathbf{p}_3+\mathbf{p}_4+\mathbf{q}-\mathbf{p}_1-\mathbf{p}_2\right).\label{eq:delta}
\end{equation}
Notice that under the mean field anzats described above, we can substitute $E_j\to \widetilde{E}_j$ in the energy-conserving delta-function as well.

\begin{figure*}
    \centering
    \includegraphics[width=0.9\linewidth]{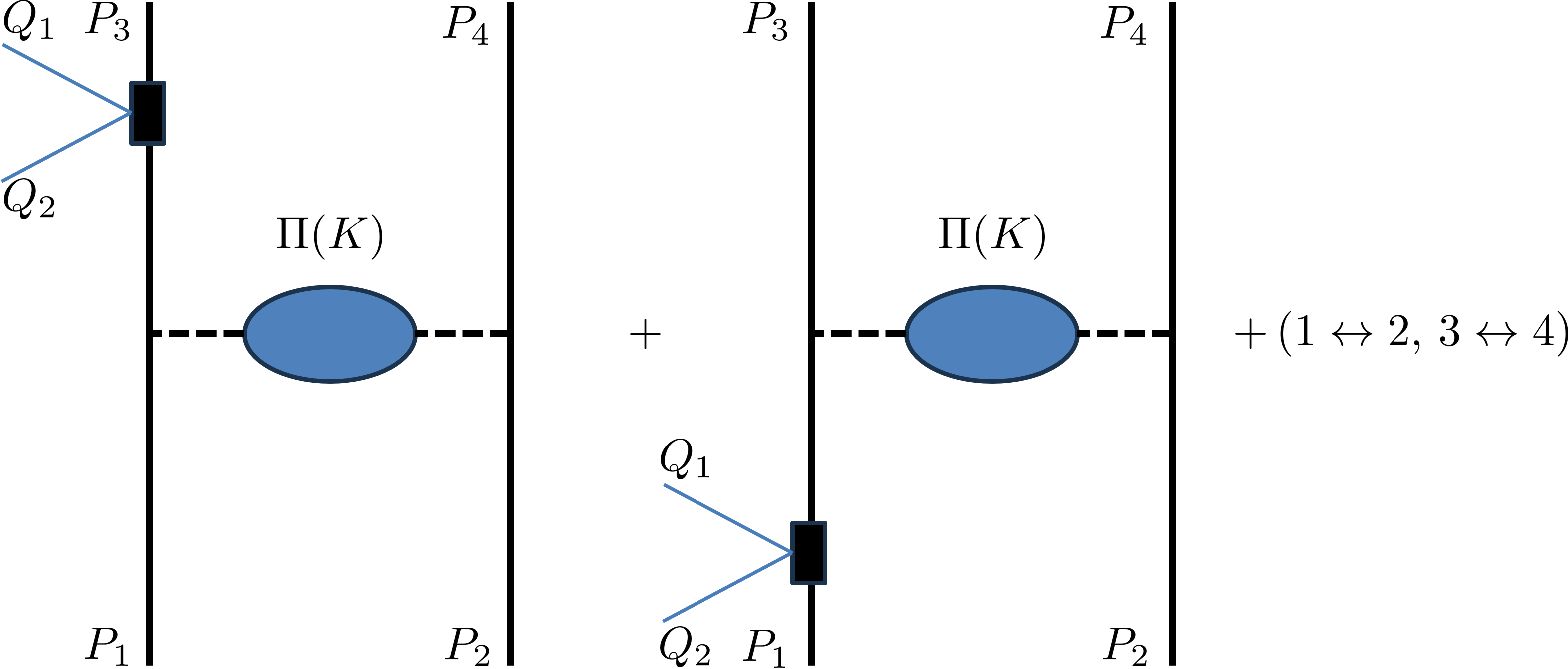}
    \caption{Feynman diagrams describing the direct part of the bremsstrahlung process amplitude. Thick solid lines refer to colliding charged fermions with four-momenta $P_{1}$, $P_2$ before and $P_3$, $P_4$ after the collision, respectively. Rectangle blocks correspond to weak vertices (in general, renormalized in medium), and thin solid lines refer to the emitted neutrino pair (with four-momenta $Q_1$ and $Q_2$). Dashed lines with bubbles correspond to the in-medium photon exchange with $\Pi(K)$ being the polarization function and $K=P_2-P_4$ being the 4-momentum transfer in a collision. An exchange contribution in the case of like particles collisions is given by an additional four diagrams obtained from the plotted ones by interchanging indices $3\longleftrightarrow 4$ (not shown in the figure).} 
    \label{fig:Amplitude}
\end{figure*}

Finally, $|M|^2$ in Eq.~(\ref{eq:Qbremms}) is the squared transition amplitude of the process, and $s$ is a symmetry factor included to compensate for double counting of the same collision events. For collisions of indistinguishable particles $s=4$ and  for collisions of different particle species $s=1$ \cite{YakovlevKaminker}.

At low temperatures, the product of Fermi-Dirac functions in Eq.~\eqref{eq:Qbremms} ensures that neutrinos are thermal, $Q_0\sim T \ll E_j$. In this case, the  soft bremsstrahlung theorem adapted for weak currents \cite{Low1958PhRv,Hanhart2001PhLB,Timmermans2002PhRvC} guarantees that the dominant contribution to emissivity is given by a set of Feynman diagrams plotted in Fig.~\ref{fig:Amplitude}, where neutrino emission is  coupled to external particle lines. The set of first-order diagrams in Fig.~\ref{fig:Amplitude} corresponds to  collisions between different particle species. In the case of indistinguishable particles, there is another set of four diagrams obtained by exchanging  the indices of the final states in Fig.~\ref{fig:Amplitude}. For long-range electromagnetic collisions, where  small values of transferred momentum are important, the direct and exchange diagrams receive dominant contributions from different kinematical domains, therefore the interference term between 
them can be omitted \cite{Kaminker1999}. The total emissivity is then just twice the contribution from the diagrams shown in Fig.~\ref{fig:Amplitude}.

Explicit expression for the amplitude graphically shown in Fig.~\ref{fig:Amplitude} is
\begin{adjustwidth}{-\extralength}{0cm}
\begin{subequations}\label{eq:M_gen}
    \begin{eqnarray}
    M &=& \frac{4\pi G_F}{\sqrt{2}}  \alpha_f l_{\alpha}\Big[ \bar{u}_3 \Gamma^{\alpha} S(P_3 + Q) \gamma^{\mu} u_1 D_{\mu \nu}(K) \bar{u}_4 \gamma^{\nu} u_2  
    +\bar{u}_3 \gamma^{\mu} S(P_1 - Q) \Gamma^{\alpha} u_1 D_{\mu \nu}(K) \bar{u}_4 \gamma^{\nu} u_2 
    \label{eq:M_first}\\
    &+& (1\leftrightarrow 2, 3 \leftrightarrow 4)\Big],\label{eq:M_second}
\end{eqnarray}
\end{subequations}
\end{adjustwidth} 
where $\alpha_f=1/137$ is the fine structure constant, $G_F =1.436\times10^{-49}$~erg~cm$^3$  ($1.166\times10^{-5}$~GeV$^{-2}$) is the Fermi weak interaction constant,  $l^{\alpha} = \bar{\psi}_{\bar{\nu}}(Q_2)\gamma^{\alpha}(1 - \gamma^5)\psi_{\nu}(Q_1)$ is the neutrino 4-current, $\psi_{\nu}$ are neutrino bispinor amplitudes, $u_j$, $j=1\dots 4$, are bispinor amplitudes of charged particles normalized as $\bar{u}_j u_j=2m_j$, and $\Gamma^{\alpha} =  \gamma^{\alpha} (c_{1V} - \gamma^5 c_{1A})$ is the effective weak interaction vertex factor in the V---A model with $c_{1V}$ and $c_{1A}$ being the vector and axial-vector form-factors for  particle species 1.
For leptons in vacuum, the form factors for the vertex $\ell\to \ell+\nu_{\ell'}+\overline{\nu}_{\ell'}$ depend on whether the emitted neutrino pair flavor $\ell'$ is the same as the emitting lepton flavor $\ell$, or not, and are
\begin{subequations}\label{eq:c_leptons}
    \begin{eqnarray}
       c_{\ell V} &=& 
       \frac{1}{2}+ 2 \sin^2\theta_W, \quad c_{\ell A} = \frac{1}{2},\quad \ell=\ell',\\
 c_{\ell V} &=& -\frac{1}{2} + 2 \sin^2\theta_W, \quad c_{\ell A} = -\frac{1}{2}, \quad \ell\neq \ell',
    \end{eqnarray}
\end{subequations}
where $\theta_W$ is the Weinberg angle ($\sin^2 \theta_W = 0.2312$). In the case of protons, emission occurs only through the neutral current channel, giving \begin{equation}\label{eq:c_protons}
    c_{pV}=(1-4\sin^2\theta_W)/2\approx 0.04, \quad c_{pA} =1.26/2 
\end{equation}
regardless of the neutrino flavors.
Above, we neglected weak magnetism contributions to the proton vector current and induced pseudoscalar contributions to the axial vector current,
as well as the momentum dependence of the proton form factors to leading order in $Q$ \cite{Timmermans2002PhRvC}. Medium effects can alter the coupling constants.
For instance, Leinson \cite{Leinson1999,Leinson2001} showed that the polarization effects renormalize the weak vertex for thermal neutrinos leading to a strong suppression of emission through the vector channel. As suggested in Ref.~\cite{Jaikumar2005PhRvD}, this correction can be included by simply dropping the vector current contribution. Following Ref.~\cite{Jaikumar2005PhRvD}, we keep here the vector current contribution for generality and set $c_{iV}\to 0$ in the final expressions. 

The function $S(P\pm Q)$ in Eq.~\eqref{eq:M_gen} is a charged fermion propagator in the internal fermion line, which we take in the form 

\begin{equation}
    S(P \pm Q) = \frac{(P \pm Q)\cdot\gamma + m_1}{(P \pm Q)^2 - m_1^{2}} \approx \pm \frac{P\cdot  \gamma + m_1}{2\ P \cdot Q}, \label{eq:S}
\end{equation}
 where $m_1$ is the quasiparticle 1 bare mass, 
and for the second equality we neglected terms of the order of $Q^2$ in the denominator. Equation~\eqref{eq:S} is a free fermion propagator. For protons within the adopted Fermi-liquid description, we assume that they obey the in-medium Dirac equation \cite{DiracBrueckner1999IRvNP} and use the same Eq.~\eqref{eq:S} but with the Dirac mass $\widetilde{m}_1$ in place of $m_1$ and `effective' energy $\widetilde{E}_j$ instead of $E_j$ in place of the time component of four-vector $P_j$ as discussed above. In this respect, we also neglect a small $\propto  T^2 \ll P\cdot Q$ quasiparticle width (imaginary part of the self-energy) in the denominator in Eq.~\eqref{eq:S}. In general, the quasiparticle propagator also contains the renormalization factor $Z$ describing the Fermi-surface depletion. Alternatively, this factor can be absorbed into the in-medium quasiparticle scattering amplitude. Here we omit this factor, noting that it can be included (in appropriate power) as a renormalization constant in the final expressions \cite{Dong2016ApJ}. Let us stress, that this approximation affects only the partial contributions involving protons to the total bremsstrahlung rates. As will be shown, when protons are in a normal state, electromagnetic bremsstrahlung is not very important, while when protons are in paired states, their contribution is  strongly suppressed.
Notice that a different approximation for the fermion propagator, namely $S(P\pm Q)=\pm Q_0^{-1}$, was used in Ref.~\cite{Jaikumar2005PhRvD} instead of Eq.~\eqref{eq:S}. However, this approximation fails, e.g., in the ultrarelativistic regime. The correct form of the propagator was used in Refs.~\cite{Kaminker1999,Jaikumar2004PhRvD}.

Finally, the term $D_{\mu \nu}(K)$ in Eq.~\eqref{eq:Qbremms} with $K\equiv(\omega,\bm{k})$ being the 4-momentum transferred in a collision event is the  in-medium photon propagator. For diagrams depicted in Fig~\ref{fig:Amplitude}, $K=P_2-P_4=P_3-P_1+Q$.  
In isotropic medium in Landau gauge, 
the photon propagator splits into the longitudinal and transverse parts via  \cite{kapusta1989finite}
\begin{eqnarray}
    D_{\mu \nu}(\omega, \bm{k}) = \frac{(\mathcal{P}_L)_{\mu \nu}}{K^2 - \Pi_L(\omega, \bm{k})}+\frac{(\mathcal{P}_T)_{\mu \nu}}{K^2 - \Pi_T(\omega, \bm{k})} , \label{eq:D_munu}
\end{eqnarray}
where $\Pi_L(\omega, \bm{k})$ and $\Pi_T(\omega, \bm{k})$ are the longitudinal and transverse polarization functions, respectively, and $\mathcal{P}_L$ and $\mathcal{P}_T$ are the longitudinal and transverse projectors, respectively. The explicit forms for the projectors are \cite{kapusta1989finite}
\begin{subequations}
    \begin{eqnarray}
        &&\mathcal{P}_T^{00} =\mathcal{ P}_T^{0 i} = \mathcal{P}_T^{i0} = 0,\\
        &&\mathcal{P}_T^{i j} = \delta^{ij} - \frac{k^i k^j}{\bm{k}^2},\\
        &&\mathcal{P}_L^{\mu \nu} = \frac{K^{\mu} K^{\nu}}{K^2} - g^{\mu \nu} - \mathcal{P}_T^{\mu \nu}.
    \end{eqnarray}
\end{subequations}

The polarization functions $\Pi_L(\omega, \bm{k})$ and $\Pi_T(\omega, \bm{k})$ describe plasma screening of the electromagnetic interaction 
and are discussed in Secs.~\ref{sec:screening_normal}--\ref{sec:screening_sc} for NS core physical conditions. They regularize the long-range (small momenta) divergence of the interaction cross-section. As will be apparent,  momentum transfer gives the dominant contribution to the emissivity.

The formalism described above applies to electromagnetic collisions between charged particles. The situation is more complicated in the case of proton-proton collisions. The proton-proton interaction is, of course, governed by strong forces, with some contribution from the long-range electromagnetic interaction. The correct way to describe proton-proton bremsstrahlung within the present formalism is to use the total amplitude $M=M_{\mathrm{em}}+M_{\mathrm{nuc}}$ in Eq.~\eqref{eq:Qbremms} that includes the nuclear contribution $M_{\mathrm{nuc}}$ and the electromagnetic contribution $M_{\mathrm{em}}$, respectively. We assume that, to a good approximation, one can neglect the interference between these two processes since the long-range $M_{\mathrm{em}}$ contributes at small transferred momenta, while all momentum values are important for $M_{\mathrm{nuc}}$. 
Within this approximation, the nuclear and electromagnetic sectors can be considered independently.

\subsection{Phasespace integrations}
Equation~\eqref{eq:Qbremms} contains  integration over an 18D space. The number of integrations can be reduced by taking the following standard steps
\begin{itemize}
    \item Introduction of the integration over the neutrino pair four momentum by inserting the identity
    \begin{equation}\label{eq:neutrino_pair_delta}
        1=\int \mathrm{d}^4 Q \delta^{(4)} (Q-Q_1-Q_2)
    \end{equation}
    into Eq.~\eqref{eq:Qbremms} and performing the integration over outgoing neutrino momenta with aid of the Lenard's identity \cite{Lenard1953PhRv}
\begin{equation}
        \int \frac{\rmd^3 q_1}{2 \omega_1} \frac{\rmd^3 q_2}{2 \omega_2} Q_{1\alpha} Q_{2\alpha'} \delta^{4}(Q - Q_1 - Q_2)
        =\frac{\pi}{24} \big(Q^2 g_{\alpha \alpha'} + 2 Q_{\alpha} Q_{\alpha'}\big) \theta(Q_0) \theta(Q^2).
\end{equation}
    This simplifies the calculation of the squared matrix element of the process, since 
    \begin{equation}
        \int \frac{\rmd^3 q_1}{2 \omega_1} \frac{\rmd^3 q_2}{2 \omega_2} \delta^{4}(Q - Q_1 - Q_2) \sum_{\mathrm{spins}}\ell_{\alpha} \ell_{\alpha'} ^*
        =\frac{4\pi}{3} \big(Q_{\alpha} Q_{\alpha'}-Q^2 g_{\alpha \alpha'} \big) \theta(Q_0) \theta(Q^2).
\end{equation}
It is then convenient to introduce spin-summed and integrated squared matrix element $\mathcal{M}$ through the expression
    \begin{equation}\label{eq:Mcal_def}
        \int \frac{\rmd^3 q_1}{2 \omega_1} \frac{\rmd^3 q_2}{2 \omega_2} \delta^{4}(Q - Q_1 - Q_2) \sum_{\mathrm{spins}}\left|M\right|^2
        =\frac{512\pi^3}{3} G_F^2 \alpha_f^2 \ \theta(Q_0) \theta(Q^2) \mathcal{M},
\end{equation}
where the normalization constant on the right-hand side is chosen for convenience. 

    \item In strongly degenerate matter, it is possible to place quasiparticles on the Fermi surfaces and separate the integrations over the angular and energy variables \cite[e.g.,][]{BaymPethick91}. One introduces dimensionless energy variables $x_i=(E_i-\mu_i)/T$ and expresses $\mathrm{d}^3 p_i =  p_{Fi}^2 v_{Fi}^{-1} T \mathrm{d}x_i \mathrm{d}\Omega_i$. 
    The integrations over $x_i$ then can be extended to the whole real axis. Inserting the integration over the transferred energy $\omega$ in a manner similar to Eq.~\eqref{eq:neutrino_pair_delta} and   taking into account thev energy-conserving delta-function in  Eq.~\eqref{eq:delta}, the integrations over $x_1,\dots,x_4$ are performed in a standard way \cite{BaymPethick91}. 

    \item 
    Since the neutrino pair momentum $q\lesssim Q_0\sim T$ is much smaller than the quasiparticle Fermi momenta and also smaller than typical values of momentum transfer $k$, we neglect it in the momentum-conserving delta function in Eq.~\eqref{eq:delta}. Then the relative orientation of the four fixed-length vectors $\bm{p}_j$ that obey the momentum conservation law is fixed by two angular variables. A convenient choice of angular variables for electromagnetic collisions is the absolute value of the transferred momentum $k$ and the angle $\phi$ between the $\{\bm{p}_1,\bm{p}_3\}$ and $\{\bm{p}_2,\bm{p}_4\}$ planes 
    \cite[e.g.,][]{Shternin2022}. In our case, when $\omega v_{Fj}\ll k$, the transferred momentum can vary between $0$ and $k_{\mathrm{max}}=2\min(p_{\mathrm{F}1},\, p_{\mathrm{F}2})$. Then, of the 8 angular integrations, only two remains:
    \begin{equation}\label{eq:electron_angles}
     \left[\prod\limits_{i = 1}^{4} \int \mathrm{d}\Omega_i \right] \delta(\bm{p}_1+\bm{p}_2-\bm{p}_3-\bm{p}_4)\ \bm{\cdot}\to \frac{8\pi^2}{p_{\mathrm{F}1}p_{\mathrm{F}2}p_{\mathrm{F}3}p_{\mathrm{F}4}}\int\limits_0^{k_{\mathrm{max}}} \mathrm{d} k\int\limits_0^{2\pi}\mathrm{d}\phi\ \bm{\cdot}.
\end{equation}\label{eq:angles}
    \item Finally, one writes $\mathrm{d}^4 Q=q^2\mathrm{d}Q_0 \mathrm{d} q \mathrm{d}\Omega_\nu$ and introduces dimensionless variables $x_\nu=q/Q_0$, $\varpi=\omega/T$, and $w=Q_0/T$. The variable $x_\nu$ runs between $0$ and $1$, the dimensionless neutrino pair energy $w$ is positive, and the dimensionless transferred energy $\varpi$  can be arbitrary.
\end{itemize}

After these steps are performed, we are left with a 7D integration. 
The general expression \eqref{eq:Qbremms} reduces to 

\begin{equation}     \label{eq:Q_brem_preint}
    Q^{\mathrm{em}}_{\mathrm{Br}} = 
    \frac{G_F^2\alpha_F^2 T^8}{24\pi^7}\
    \int\limits_{-\infty}^{+\infty} \mathrm{d}\varpi \int\limits_0^\infty \mathrm{d}w\  
     \mathcal{Z}(w,\varpi) \int\limits_0^{k_{\mathrm{max}}} \mathrm{d} k\, \frac{\left\langle\mathcal{M}\right\rangle_{\nu\phi}}{s},
\end{equation}
where the energy kernel is 
\begin{equation}\label{eq:Zfac}
     \mathcal{Z}(w,\varpi)=\frac{w^4 \varpi (\varpi +w)}{(1-\mathrm{e}^{-\varpi})(\mathrm{e}^{w+\varpi}-1)} 
\end{equation}
and we combined the integrations over $\phi$ and neutrino pair variables $\Omega_\nu$ and $x_\nu$ 
into the term
\begin{equation}     \label{eq:Mnuphi}
\left\langle\mathcal{M}\right\rangle_{\nu\phi}=
     \int\limits_0^1 x_\nu^2 \mathrm{d}x_\nu\int \frac{\mathrm{d}\Omega_\nu}{4\pi}\int\limits_0^{2\pi}\frac{\mathrm{d}\phi}{2\pi} \ 
     \frac{\mathcal{M}}{s}.
\end{equation}
The reason for such a combination will be evident in a moment. 

The spin-summed squared matrix element $\mathcal{M}$ in Eq.~\eqref{eq:Mnuphi} 
can be written as a sum of three terms
\begin{equation}    
\mathcal{M}\equiv\sum_{\mathrm{spins}} |M|^2 =\mathcal{M}_{12} + \mathcal{M}_{21}+\mathcal{M}_{\mathrm{inf}},
\end{equation}
where $\mathcal{M}_{12}$ corresponds to the squared amplitude of the process where the particle $1$ emits neutrino (either in the initial state 1 or in the final state $3$) while scattering off particle 2. That is, this is the process shown explicitly in Fig.~\ref{fig:Amplitude} and whose amplitude is given in Eq.~\eqref{eq:M_first}. The term $\mathcal{M}_{21}$  results from a similar process where now particle 2 emits while being scattered off particle 1 (the respective amplitude is encapsulated in the $(1\leftrightarrow 2, 3 \leftrightarrow 4)$ term in Fig.~\ref{fig:Amplitude} and Eq.~\eqref{eq:M_second}), and $\mathcal{M}_{\mathrm{inf}}$ is the interference contribution between these two processes. The calculations show  (see also Ref.~\cite{Kaminker1999}) that the
$\mathcal{M}_{\mathrm{inf}}$ contribution is small and can be neglected. Remind that we also neglected the interference term between the direct and exchange contributions in the case of collisions of like particles, leading to a doubling of $\mathcal{M}$. Furthemore, for like particles $\mathcal{M}_{12}=\mathcal{M}_{21}$. Thus, in a multicomponent mixture one can substitute
\begin{equation}
    \frac{\mathcal{M}}{s}\to\mathcal{M}_{12}
\end{equation}
in Eq.~(\ref{eq:Mnuphi}) and sum the final bremmstrahlung emissivity over all ordered pairs of particles, i.e.
\begin{equation}\label{eq:Qbr_sum}
Q_{\mathrm{Br}}^{\mathrm{tot}} = \sum_{ij} Q_{ij}.
\end{equation}

Performing the necessary spin summations, working out the contraction with neutrino tensor, and putting  $Q\to0$ where possible, we obtain
\begin{subequations}\label{eq:M12gen}
\begin{eqnarray}
\mathcal{M}_{12}&=&S_{13}^{\mu\mu'}T_{24}^{\nu\nu'} D_{\mu\nu}(P_2-P_4)D_{\mu'\nu'}^*(P_2-P_4),\label{eq:M12}
\end{eqnarray}
where
\begin{adjustwidth}{-\extralength}{0cm}
\begin{eqnarray}
   S_{13}^{\mu\mu'}&=&\frac{8 Q^2}{uv} \big(P_1\cdot P_3-m_1^{2}\big)\Bigg[\big(c_{1V}^2+c_{1A}^2\big)T_{31}^{\mu\mu'}+2ic_{1V}c_{1A} \epsilon^{P_1P_3\mu\mu'} + 2m_1^{2} c_{1A}^2\Big(\frac{Q^\mu Q^{\mu'}}{Q^2} -2g^{\mu\mu'}\Big)\Bigg]\label{eq:S13line1}\\
    &&+4Q^2m_{1}^2\Big(\frac{1}{u}-\frac{1}{v}\Big)^{2}\Big(2c_{1A}^2-c_{1V}^2\Big)T_{31}^{\mu\mu'} -2m_1^{2} c_{1A}^2 \frac{(u-v)^2}{uv} g^{\mu\mu'}\label{eq:S13line2}\\
    &&-\frac{16m_{1}^{2} c_{1A}^2 Q^2}{uv} (P_1-P_3)^\mu(P_1-P_3)^{\mu'} -\frac{4m_1^{2}c_{1A}^2(u-v)}{uv} (P_1-P_3)^{(\mu}Q^{\mu')},\label{eq:S13line3}\\
    T_{31}^{\mu\mu'} &=&  \frac{1}{4}\mathrm{Sp}\big[(P_{1} + m_{1}) \gamma^{\mu}(P_{3} + m_{1}) \gamma^{\mu'}\big]
    =P_3^{(\mu}P_1^{\mu')} + \big(m_1^{2}-P_1\cdot P_3\big) g^{\mu\mu'},
\end{eqnarray}
\end{adjustwidth}
\end{subequations}
and $T^{\nu\nu'}_{24}$ is obtained from  $T^{\nu\nu'}_{31}$ by substituting $P_1\to P_2$, $P_3\to P_4$, and $m_1\to m_2$. In Eq.~\eqref{eq:M12gen} we abbreviated $u=2P_3\cdot Q$, $v=2P_1\cdot Q$, and used a  the shorthand notation for the symmetrized product $P_3^{(\mu}P_1^{\mu')}=P_3^{\mu}P_1^{\mu'}+P_1^{\mu}P_3^{\mu'}$. In ultrarelativistic case relevant for electrons (and to a large extent for muons), the contributions given in lines \eqref{eq:S13line2}--\eqref{eq:S13line3} as well as the third term in brackets in line \eqref{eq:S13line1} vanish; therefore the expression simplifies considerably.
We checked, that using the photon propagator of Ref.~\cite{Kaminker1999} in Eq.~\eqref{eq:M12} and the ultrarelativistic approximation for electrons we reproduce their results for the direct part of the amplitude squared corresponding to $ee$ and $ep$ bremsstrahlung. We are not reproducing the results of Ref.~\cite{Jaikumar2004PhRvD} in the ultrarelativistic case because of a different expression for the electron propagator. 

 Now we substitute the explicit form of the photon propagator \eqref{eq:D_munu} into Eq.~\eqref{eq:M12} and perform the integrations in Eq.~\eqref{eq:Mnuphi}. It turns out, that these integrations can be performed analytically. The structure of Eq.~\eqref{eq:M12} suggests that, in principle, the resulting expression should contain the terms with products of the longitudinal and transverse parts of the propagator. However, these terms vanish after integration over $\phi$ (see Appendix~\ref{app:MandF} for details). For the same reason, the mixed term between the vector and axial-vector channels [see Eq.~\eqref{eq:S13line1}] also vanishes. As a result, the final expression contains four terms corresponding to contributions of the longitudinal and transverse parts of the electromagnetic interaction to the vector and axial-vector neutrino emission channels and can be written as
\begin{equation} \label{eq:M12nuphi_calc}    \left\langle\mathcal{M}_{12}\right\rangle_{\nu\phi}=
\sum_{r=V,A} c_{1r}^2k^2\left[\frac{{4E_{\mathrm{F}2}^2-k^2}}{|k^2+\Pi_L(\omega, k)|^2} F_{Lr}(v_{\mathrm{F}1},\kappa_1)
+\frac{{4p_{F2}^2+k^2}}{|k^2+\Pi_T(\omega,k)|^2} F_{Tr}(v_{\mathrm{F}1},\kappa_1)
\right],
\end{equation}
where the four factors $F_{LV}$, $F_{LA}$, $F_{TV}$, and $F_{TA}$ depend on the particle 1 Fermi velocity $v_{\mathrm{F}1}$ and the parameter $\kappa_1=k/(2p_{\mathrm{F}1})$. In Eq.~\eqref{eq:M12nuphi_calc} we used the fact that $\omega\ll k$ hence we approximate $K^2\approx -k^2$ in denominators. 
Since vector and axial-vector channels enter Eq.~\eqref{eq:M12nuphi_calc} incoherently, the summation over neutrino flavors can be performed. This amounts to replacement of $c^2_{1r}$ with the total $C_{1r}^2\equiv \sum_\nu c_{1r}^2(\nu)$ in 
Eq.~\eqref{eq:M12nuphi_calc} and subsequent expressions.

Full explicit expressions for the four factors $F_{L/T r}$ are cumbersome and given in Appendix~\ref{app:MandF}. Non-relativistic ($v_{\mathrm{F}1}\to 0$) and ultrarelativistic ($v_{\mathrm{F}1}\to 1$)  limits for these factors are collected in Table~\ref{tab: Flims}. We note that only the $F_{TV}$ contribution is actually suppressed in the non-relativistic limit. Since small momentum transfers dominate, in practice the leading order expressions in $k$ are needed. They are obtained by  setting $\kappa_1=0$ in expressions for the factors $F$, i.e. $F^{lo}_{LV}(v_{\mathrm{F}1})\equiv F_{LV}(v_{\mathrm{F}1},0)$ and similarly for the three remaining factors. Explicit expressions  
read
\begin{adjustwidth}{-\extralength}{0cm}
\begin{eqnarray}
F_{LV}^{lo} (v_{\mathrm{F}1}) &=& \frac{15-v_{\mathrm{F}1}^2-8v_{\mathrm{F}1}^4}{18v_{\mathrm{F}1}^6} +\frac{-5+2v_{\mathrm{F}1}^2+3v_{\mathrm{F}1}^4}{6 v_{\mathrm{F}1}^7}\,\mathrm{ArcTanh}\,v_{\mathrm{F}1},\label{eq:FLV_lo}
\\
F_{TV}^{lo}(v_{\mathrm{F}1})&=&\frac{1}{2}v_{\mathrm{F}1}^2F_{LV}^{lo}(v_{\mathrm{F}1}),\label{eq:FTV_lo}\\
F_{LA}^{lo}(v_{\mathrm{F}1}) &=& \frac{-30+11v_{\mathrm{F}1}^2+37 v_{\mathrm{F}1}^4-12v_{\mathrm{F}1}^6}{18v_{\mathrm{F}1}^6} +\frac{10-7v_{\mathrm{F}1}^2-12v_{\mathrm{F}1}^4+9v_{\mathrm{F}1}^6}{6 v_{\mathrm{F}1}^7}\,\mathrm{ArcTanh}\,v_{\mathrm{F}1},\label{eq:FLA_lo}
\\
F_{TA}^{lo}(v_{\mathrm{F}1})&=& \frac{33-67v_{\mathrm{F}1}^2+40 v_{\mathrm{F}1}^4}{36v_{\mathrm{F}1}^4} -\frac{11-26v_{\mathrm{F}1}^2+15v_{\mathrm{F}1}^4}{12 v_{\mathrm{F}1}^5}\,\mathrm{ArcTanh}\,v_{\mathrm{F}1}.\label{eq:FTA_lo}
\end{eqnarray}
\end{adjustwidth}
In the ultrarelativistic case, these factors simply reduce to $F_{LV}^{lo}=F_{LA}^{lo}=2F_{TV}^{lo}=2F_{TA}^{lo}=1/3$.
\renewcommand\arraystretch{1.5}
\begin{table}[H] 
\caption{Limiting expressions for factors $F_{Lr}$ and $F_{Tr}$, $r=V,A$ in Eq.~\eqref{eq:M12nuphi_calc}. The rows marked `NR' and `UR' correspond to the non-relativistic and ultrarelativistic limits, respectively.\label{tab: Flims}}
\begin{adjustwidth}{-\extralength}{0cm}
\begin{tabularx}{\fulllength}{CCCCC}
\toprule
&$F_{LV}$	& $F_{LA}$	& $F_{TV}$& $F_{TA}$\\
\midrule
NR		& $\frac{6}{35}$		& $\frac{4}{35}$	 & $\frac{2}{35}v_{\mathrm{F}1}^2(1+\kappa_1^2)$& $\frac{2}{5}$\\
UR		& $\frac{1-\kappa_1^2+2\kappa_1^2\log\kappa_1}{3(1-\kappa_1^2)}$			&$\frac{1-\kappa_1^2+2\kappa_1^2\log\kappa_1}{3(1-\kappa_1^2)}$	& $\frac{(1-\kappa_1^2+2\kappa_1^2\log\kappa_1)(1+\kappa_1^2)}{6(1-\kappa_1^2)^2}$&$\frac{(1-\kappa_1^2+2\kappa_1^2\log\kappa_1)(1+\kappa_1^2)}{6(1-\kappa_1^2)^2}$	\\
\bottomrule
\end{tabularx}
\end{adjustwidth}
\end{table}

Inserting Eq.~\eqref{eq:M12nuphi_calc} into Eq.~\eqref{eq:Q_brem_preint} and summing over the neutrino flavors, we obtain the final expression for the bremmstrahlung neutrino pair  emissivity of a  particle $1$  scattered off particle 2 
\begin{equation}\label{eq:Q_12noscr}
    Q_{12}=\frac{G_F^2 \alpha_f^2 T^8}{24\pi^7} 
    \sum_{r=V,A} C_{1r}^2 \left[\mathcal{F}_{Lr}+\mathcal{F}_{Tr}\right],
\end{equation}
where 
\begin{equation}\label{eq:FintLr}
    \mathcal{F}_{Lr} =
    \int\limits_{-\infty}^{+\infty} \mathrm{d}\varpi \int\limits_0^\infty \mathrm{d}w\  
    \mathcal{Z}(w,\varpi)
     \int\limits_0^{k_{\mathrm{max}}} \mathrm{d} k\ k^2\frac{{4E_{\mathrm{F}2}^2-k^2}}{|k^2+\Pi_L(\omega, k)|^2} F_{Lr} (v_{\mathrm{F}1},\kappa_1) 
\end{equation}
and
\begin{equation}\label{eq:FintTr}
    \mathcal{F}_{Tr} =
    \int\limits_{-\infty}^{+\infty} \mathrm{d}\varpi \int\limits_0^\infty \mathrm{d}w\  
    \mathcal{Z}(w,\varpi)
     \int\limits_0^{k_{\mathrm{max}}} \mathrm{d} k\ k^2\frac{{4p_{\mathrm{F}2}^2+k^2}}{|k^2+\Pi_T(\omega, k)|^2} F_{Tr}(v_{\mathrm{F}1},\kappa_1).
\end{equation}

In order to evaluate the remaining integrals, the  plasma screening needs to be specified, which is different in normal and superconducting matter. This is done in the next two subsections \ref{sec:screening_normal} and \ref{sec:screening_sc}. Before turning to the description of the plasma screening, it is possible to make some general remarks on the obtained expressions. The structure of the result in Eqs.~\eqref{eq:Q_12noscr}--\eqref{eq:FintTr} resembles, for natural reasons, the expressions obtained when one calculates the transport coefficients mediated by the same collision events \cite[e.g.,][]{SchmittShternin2018,Shternin2022}. The quantities $\mathcal{F}_{Lr}$ and $\mathcal{F}_{Tr}$ can be viewed as effective Coulomb  logarithms of the bremsstrahlung problem, although there is no actual logarithm under the NS core conditions since the temperature is much smaller than the plasma temperature there. As already stated, small values of $k$ (aka the small-angle approximation) dominate the innermost integrals in Eqs.~\eqref{eq:FintLr}--\eqref{eq:FintTr}. In the leading order, the integrands are proportional to $k^2$ in numerators, which is explicitly emphasized in Eq.~\eqref{eq:M12nuphi_calc} and Eqs.~\eqref{eq:FintLr}--\eqref{eq:FintTr}. In transport coefficients studies, a similar structure of collision frequencies is obtained for the shear viscosity or electrical conductivity/diffusion problem \cite[e.g.,][]{SchmittShternin2018,Shternin2022}. In contrast, Jaikumar et al.~\cite{Jaikumar2005PhRvD} obtained expressions for effective collision frequencies that have the same general structure, but with the leading order of $k^0$ in the numerators. Such collision frequencies are encountered in studies of the thermal conductivity problem and are considerably larger in dense matter. 
Indeed, the absence of the small $k^2$ factor severely enhances the collisions rate and thus the bremsstrahlung emissivity (see also discussion in Ref.~\cite{Kaminker1999}). 
The absence of $k^2$ in the expressions of Ref.~\cite{Jaikumar2005PhRvD} can be traced to the obscure selection of $\pm Q_0^{-1}$  as the electron propagator in that work instead of the correct form \eqref{eq:S}.

\subsection{Screening in normal matter}\label{sec:screening_normal}
We start with normal (i.e. not superconducting) matter. 
We are concerned with small energy and momentum transfer, i.e. ($\omega,\ k \ll p_{\mathrm{F}j}$). Moreover, the transferred energy is of the order of temperature, thus one has $\omega v_{\mathrm{F}j}\ll k$. Therefore, for the NS core conditions it is enough to consider screening in the hard dense loop and static limit \cite{Shternin2022}. In this limit, the longitudinal polarization function reduces to  the Thomas-Fermi screening:
\begin{equation}\label{eq:Pil_norm}
\Pi_L(\omega,k)\approx\Pi_L(0,k)=q_{\mathrm{TF}}^2\equiv \frac{4\alpha_f}{\pi}\sum_{j=e,p,\mu}  \frac{p_{\mathrm{F}j}^2}{v_{\mathrm{F}j}},
\end{equation}
where $q_{\mathrm{TF}}$ is the characteristic Thomas-Fermi momentum. From Eq.~\eqref{eq:Pil_norm} and Eq.~\eqref{eq:FintLr}, it is clear that the momentum transfers $k\lesssim q_{\mathrm{TF}}$ dominate the collisions in the longitudinal sector. 
In contrast, the transverse screening vanishes in the static limit. In the leading order in $\omega/k$, the transverse polarization function is 
\begin{equation}\label{eq:Pit_norm}
    \Pi_T(\omega,k)=
    i\frac{\omega}{k} \alpha_f\sum_{j=e,p,\mu}   p_{\mathrm{F}j}^2\equiv i\Lambda^3/k,
\end{equation}
where $\Lambda\propto \omega^{1/3}$ is the characteristic  dynamic momentum transfer scale in the transverse sector. Since $\omega\sim T$, one typically has $\Lambda\ll q_{\mathrm{TF}}$ therefore the transverse part of the interaction has a considerably larger range than the longitudinal part of the interaction, providing the dominant contribution to the electromagnetic collision rates in relativistic degenerate matter \cite{Baymetal1990,Heiselberg1992,HeiselbergPethick1993PhRvD,SchmittShternin2018,Shternin2022}. 

Nevertheless, let us start with the longitudinal contribution to Eq.~\eqref{eq:Q_12noscr}. Since the screening in Eq.~\eqref{eq:Pil_norm} is static, the integration over the energy variables $w$ and $\varpi$ can be performed independently, giving
\begin{equation}\label{eq:zeta0}
    \int\limits_{-\infty}^{+\infty} \mathrm{d}\varpi \int\limits_0^\infty \mathrm{d}w\  
     \mathcal{Z}(w,\varpi) \equiv\zeta_0 \approx 1647.7
\end{equation}
and only one integration over $k$ remains in Eq.~\eqref{eq:FintLr}. Due to the complex form of the factors $F_{Lr}(v_{\mathrm{F}1},\kappa_1)$, this integration cannot be performed analytically and should be done numerically. However, we can rely on the small angle approximation, neglecting terms of higher order in $k^2$.  In this approximation, 
\begin{equation}\label{eq:FintLr_lo}
    \mathcal{F}_{Lr}\approx\mathcal{F}^{lo}_{Lr}=\zeta_0 
    \frac{4p_{\mathrm{F}2}^2}{v_{\mathrm{F}2}^2q_{\mathrm{TF}}}F_{Lr}^{lo}(v_{\mathrm{F}1}) I_L^{(2)}(k_{\mathrm{max}}/q_{\mathrm{TF}}),
\end{equation}
where
\begin{equation}\label{eq:IL2_lo}
    I_L^{(2)}(y_m)=\int\limits_0^{y_m} \frac{y^2\ \mathrm{d} y}{(y^2+1)^2} = \frac{1}{2}\left(\mathrm{ArcTan} \ y_m - \frac{y_m}{y_m^2+1} \right).
\end{equation}
Asymptotically, $I_L^{(2)}(y_m)\to\pi/4$ for $y_m \to \infty$, and $I_L^{(2)}(y_m)\sim y_m^3/3$ for $y_m \to 0$. At first glance, the small values of $k_{\mathrm{max}}/q_{\mathrm{TF}}$ are outside the small-angle approximation. However, when the Fermi momentum of the target particle $p_{\mathrm{F}2}$ is very small, which  in NS core conditions happens for muons near the threshold of their appearance, one still has $\kappa_1 \ll 1$, $k^2\ll 4E_2^2$, but $k_{\mathrm{max}}=2p_{\mathrm{F}2}\lesssim q_{\mathrm{TF}}$. Strictly speaking, in this case the approximation \eqref{eq:Pil_norm} for muons becomes invalid, but they do not contribute to the screening in this limit anyway.
Therefore, the general form \eqref{eq:FintLr_lo} properly describes scattering off the target particles. On the other hand, it fails when the density of the emitting particle species $1$ is close to the threshold, so that the condition $q_{\mathrm{TF}}\ll p_{\mathrm{F}1}$ no longer holds.

In the transverse sector, the screening is dynamic and depends on $\omega$. Therefore, the integration in Eq.~\eqref{eq:FintTr} cannot be performed in advance. However, due to the very small value of the characteristic screening momentum $\Lambda$, the weak screening approximation holds in almost any case of practical interest. By making use of
\begin{equation}
\int\limits_0^{k_{\mathrm{max}}} \frac{\mathrm{d} k\ k^2}{|k^2+i\Lambda^3 /k|^2} \approx \frac{\pi }{3|\Lambda|},\quad |\Lambda|\ll k_{\mathrm{max}},
\end{equation}
we obtain for the lowest order transverse contribution
\begin{equation}\label{eq:FTr_int_lo}
    \mathcal{F}_{Tr}^{lo}=
    \zeta_{-1/3} \frac{4\pi p_{\mathrm{F}2}^2}{3\Lambda_T}F_{Tr}^{lo}(v_{\mathrm{F}1}),
\end{equation}
where $\Lambda_T$ is defined via $\Lambda=\varpi^{1/3}\Lambda_T$, i.e.
\begin{equation}\label{eq:LambdaTdef}
    \Lambda_T=\left(T \alpha_f\sum_j p_{\mathrm{F}j}^2\right)^{1/3},
\end{equation}
and the constant $\zeta_{-1/3}$  results from the energy integration
\begin{equation} \label{eq:zeta1/3}
    \zeta_{-1/3}\equiv \int\limits_{-\infty}^{+\infty} \mathrm{d}\varpi \int\limits_0^\infty \mathrm{d}w\  
     \mathcal{Z}(w,\varpi)|\varpi|^{-1/3}\approx 1278.
\end{equation}
Details on numerical calculations beyond the small-angle approximation in the transverse channel are given in Appendix~\ref{app:num_trans}.

The dynamical character of plasma screening in the transverse channel leads to a peculiar temperature dependence of the corresponding contribution to the bremsstrahlung emissivity. Since $\Lambda_T$ enters Eq.~\eqref{eq:FTr_int_lo} in the denominator and, according to Eq.~\eqref{eq:LambdaTdef}, is very small, being proportional to $T^{1/3}$, the transverse part of the electromagnetic interaction dominates the bremsstrahlung emissivity. To leading order, $Q^{\mathrm{em}}_{\mathrm{Br}}$ therefore has  a non-standard temperature dependence $Q^{\mathrm{em}}_{\mathrm{Br}}\propto T^{23/3}$ instead of $Q^{\mathrm{em}}_{\mathrm{Br}}\propto T^{8}$. Notice that the dominance of the transverse contribution to neutrino pair electromagnetic bremsstrahlung emissivity was first pointed out by  Jaikumar et al.~\cite{Jaikumar2005PhRvD}.
However, they obtained a much stronger increase in the emissivity in the transverse channel because of the specific selection of the 
fermion propagator. Due to the lack of a $k^2$ factor in their expressions, the effective collision frequencies were found to be much higher, and proportional to $\Lambda_T^{-3}\propto T^{-1}$ instead of $\Lambda_T^{-1}\propto T^{-1/3}$ here. As a result, neutrino emissivities in the electron-electron bremsstrahlung process in Ref.~\cite{Jaikumar2005PhRvD} were found to be comparable to or larger than some of the nucleon bremsstrahlung processes. Here we obtain much more modest values of the emissivity. We can conclude (see discussion in Sec.~\ref{sec:discus} below) that the calculation of neutrino electromagnetic bremsstrahlung in the absence of nucleon pairing is of academic interest only.

\subsection{Screening in superconducting matter}\label{sec:screening_sc}

The electromagnetic bremshtrahlung neutrino emission process considered here is only relevant when all standard emission mechanisms in the baryon sector are strongly suppressed by pairing. In particular, this means that it is enough in practice to consider the case of a well-developed proton superconductivity, $T\ll T_{Cp}$.
In this limit, the contribution from protons to the neutrino emission in electromagnetic bremsstrahlung is suppressed as well; therefore it is necessary to consider the lepton sector only. We are closely following Ref.~\cite{Shternin2018}, devoted to lepton transport coefficients in a similar regime, and refer to that work for additional details. 

It is believed that proton pairing in NS cores occurs in $^1S_0$ state with a density-dependent critical temperature \cite[e.g.,][]{LombardoSchulze2001LNP,SedrakianClark2019}. It leads to the appearance of an isotropic  energy gap $\Delta$ in the spectrum of single-particle excitations  of the proton liquid. The appearance of proton pairing leads to modification of the screening of electromagnetic interactions. Up to order of $\Delta/\mu_p$, the static longitudinal screening does not change \cite{Gusakov2010PhRvC, Arseev2006PhyU} and $\Pi_L(\omega,\bm{k})$ is given by Eq.~(\ref{eq:Pil_norm}). This means that the longitudinal contribution to the emissivity, Eq.~\eqref{eq:Q_12noscr}, remains intact and Eqs.~\eqref{eq:FintLr} or \eqref{eq:FintLr_lo} can be used in the case of superconducting protons  as well.
 
In contrast, the transverse screening modifies. It becomes static and dominated by protons.  In the limit of low temperatures, $T\to 0$, within the Bardeen-Cooper-Schrieffer (BCS) theory, the transverse polarization function  is \cite{Landau9eng}
\begin{equation}\label{eq:Pit_static}
\Pi_T(\omega,\bm{k})\approx\Pi_{Tp}(0,k)=q_M^2 J(\zeta),
\end{equation}
where $\zeta= k v_{\mathrm{F}p}/\Delta$ 
and $q_M$ is the Meissner mass [cf. Eqs.~\eqref{eq:Pil_norm}--\eqref{eq:Pit_norm}]
\begin{equation}\label{eq:Meissner}
 q_M^2=\frac{4\alpha
_f}{3\pi} p_{\mathrm{F}p}^2 v_{\mathrm{F}p}.
\end{equation}
The function $J(\zeta)$ in Eq.~(\ref{eq:Pit_static}) is \cite{Landau9eng}
\begin{eqnarray}\label{eq:J_T0}
J(\zeta)&=&\frac{3}{2}\int\limits_{-1}^{1} d x\, (1-x^2)\, \frac{{\rm ArcSinh}(\zeta x/2)}{\zeta x \sqrt{1+(\zeta x)^2/4}}
\end{eqnarray}
and has asymptotic behavior $J(0)=1$ and $J(\zeta)\sim 3\pi^2/(4\zeta)$ at $\zeta \to \infty$. In the so-called London limit $k v_{\mathrm{F}p} \ll \Delta$ (small $\zeta$), the screening in Eq.~\eqref{eq:Pit_static}
is therefore given by the Meissner mass $q_M$ and does not depend on $k$ or $\Delta$. In the opposite Pippard limit $k v_{\mathrm{F}p} \gg \Delta$, the polarization function \eqref{eq:Pit_static} is proportional to $\Delta$ and inversely proportional to $k$. This behavior resembles in a certain way Eq.~(\ref{eq:Pit_norm}) with $\Delta$ in place of $\omega$. Inspection of Eq.~\eqref{eq:FintTr} reveals that which of the limits discussed above is important in practice is described by the parameter $A=\zeta(k=q_{M})=q_Mv_{\mathrm{F}p}/\Delta$ \cite{Shternin2018}. For $A\ll 1$, the dominant contribution to the integral in Eq.~\eqref{eq:FintTr} comes from small values of $k$ and the London expression $\Pi_{Tp}(0,k)=q_M^2$ can be used. This approximation was briefly discussed in Ref.~\cite{Jaikumar2005PhRvD}. It fails if  $A$ happens to be much greater than unity, where the Pippard limit is more appropriate. For a fixed $\Delta$, the parameter $A$ increases with density. Investigation shows, that for reasonable values of the proton superconductivity gap, the parameter $A$ in NS cores can vary over a wide range between 0.01 and 100 \cite{Shternin2018}. This means that one cannot rely on certain asymptotic results and should consider the general case. 
Notice that the value $A=\sqrt{2}\pi$ corresponds to the transition from a type-II superconductor at low densities in the outer NS core to a type-I superconductor at higher densities, see Ref.~\cite{Shternin2018} for details.  

Because the screening is static, the energy integral in Eq.~\eqref{eq:FintTr} can be easily performed, giving $\zeta_0$, Eq.~\eqref{eq:zeta0}, as in the longitudinal case. Moreover, since the characteristic screening scale $q_M$ in Eq.~\eqref{eq:Pit_static} is small (it is always $J(\zeta)<1$), we can still use the small-angle approximation. In this limit, Eq.~\eqref{eq:FintTr} reduces to [cf. Eq.~\eqref{eq:FintLr_lo}]
\begin{equation}\label{eq:FintTr_SF_lo}
\mathcal{F}^{SF,lo}_{Tr}=\zeta_0 
    \frac{4p_{\mathrm{F}2}^2}{q_{M}}F_{Tr}^{lo}(v_{\mathrm{F}1}) I_{TSF}^{(2)}(k_{\mathrm{max}}/q_{M},A),
\end{equation}
where 
\begin{equation}\label{eq:ITSF2_lo}
    I_{TSF}^{(2)}(y_m,A)=\int\limits_0^{y_m} \frac{y^2\ \mathrm{d} y}{(y^2+J(Ay))^2}.
\end{equation}
The function $I_{TSF}^{(2)}(y_m,A)$ can be analyzed in various limiting cases. In the London limit, $A\ll 1$, it is given by Eq.~\eqref{eq:IL2_lo}. The opposite, Pippard limit realizes when $A\gg 1$, but also $A y_m\gg 1$. In this case
    \begin{equation}
       I_{TSF}^{(2),Pip}(y_m,A) = \left(\frac{4A}{3\pi^2}\right)^{1/3}  I_{Pip}\left(y_m \left[4A/(3\pi^2)\right]^{1/3}\right),
    \end{equation}
    where   \begin{equation}\label{eq:ITSF_2_Pip}
       I_{Pip}(\xi) = \int\limits_0^\xi \frac{t^4\ \mathrm{d}t}{(t^3+1)^2}=\frac{1}{27}\left(\sqrt{3}\pi +6\sqrt{3}\,\mathrm{ArcTan}\frac{2 \xi-1}{\sqrt{3}}-\frac{9\xi^2}{1+\xi^3} +3\ln \frac{\xi^2-\xi+1}{(\xi+1)^2}\right).
    \end{equation}
However, the most important limit in practice corresponds to the case of large $y_m$:
\begin{equation}\label{eq:ITSF2_yminf}
    I_{TSF}^{(2)}(y_m,A) = f_2(A)-\frac{1}{y_m} + o\left(y_m^{-1}\right),\quad y_m\to \infty,
\end{equation}
where $f_2(A)$ is given by Eq.~\eqref{eq:ITSF2_lo} with the upper integration limit set to $\infty$. Asymptotically, $f_2(0)=\pi/4$ and $f_2(A)\sim 4\pi/(9\sqrt{3}) (4A/3\pi^2)^{1/3}$ at $A\to \infty$.
    The fitting expression which respects both  asymptotics is      \begin{equation}\label{eq:f2A}
f_2(A)=\left(0.269+0.0045\  A^{1.21}+0.0082\  A^{1.81}\right)^{0.184}.
    \end{equation} 
The relative approximation error of Eq.~\eqref{eq:f2A} in the range $10^{-5}<A<10^5$ does not exceed 0.6\%. 

The limit of small or intermediate $y_m$ can be relevant to describe scattering off muons near the threshold (see the discussion in the previous section). For very small $y_m\ll 1$ and $y_m A\ll 1$, $I_{TSF}^{(2)}\sim y_m^3/3$. For intermediate $y_m>1$, a fitting expression for $I_{TSF}^{(2)}(y_m,A)$ was constructed in Ref.~\cite{Shternin2018}. Here we provide an updated fit, which is valid for any values of $y_m$ and $A$ and respects the limiting cases:
\begin{subequations}\label{eq:ITSF_2_fit}
    \begin{equation}
    I_{TSF}^{(2)}(y_m,A) = \frac{9\sqrt{3}}{4\pi}f_2(A_{\mathrm{eff}}) I_{Pip}\left(y_m \left[4A_{\mathrm{eff}}/(3\pi^2)\right]^{1/3}\right),
\end{equation}
where
\begin{equation}
   A_{\mathrm{eff}}(y_m,A)=\left(A^{p_0}+B(y_m)^{p_0}\right)^{1/p_0}, \end{equation}
\begin{equation}
    B(y_m)=\frac{\sqrt{15}\pi^2}{4y_m}\left(1+b_1y_m^{p_1}+b_2 y_m^{p_2}\right)^{p_3},
\end{equation}
\end{subequations}
and the fit parameters $p_0,\dots,p_3$, $b_1$, and $b_2$ are given in Table~\ref{tab:fit_params}. The typical relative error for this expression is less than 1\%; however, at small $y_m$, the maximal relative error of the fit is about 10\%.

\begin{table}[H] 
\caption{Fitting parameters in Eq.~\eqref{eq:ITSF_2_fit} }\label{tab:fit_params}
\begin{tabularx}{\textwidth}
{CCCCCC}
\toprule
$p_0$& $p_1$ & $p_2$ & $p_3$ & $b_1$ & $b_2$\\
\midrule
1.206 & 0.725& 1.249& 0.3432& $-0.5802$& 0.6347\\
\bottomrule
\end{tabularx}
\end{table}

\section{Results and discussion}\label{sec:discus}
For illustration, we select the modern realistic equation of state (EoS) BSk24 \cite{BSk2018}. This EoS is based on the family of Brussels-Skyrme density functionals and belongs to the class of so-called unified EoSs, which can describe the matter in NS crust and the core based on the same principles. The BSk24 EoS satisfies most of the current astrophysical and nuclear experiments constraints \cite[e.g.,][]{OfengeimEncycl2026}. Nevertheless, this EoS is selected for illustrative purposes, and the results would be qualitatively the same for any realistic  nucleon EoS. For this EoS, muons appear at a density $\rho=0.76\rho_0$, where $\rho_
0=2.8\times 10^{14}$~g~cm$^{-3}$ is the nuclear saturation density,  and the maximal mass reached in BSk24 NSs is $M=2.28M_\odot$ which corresponds to the central density of $\rho=8.1\rho_0$.
We do not use, however, the proton effective masses from the BSk24 EoS.
Instead, we use the constant $\widetilde{m}_p=0.8 m_N$, where $m_N=939$~MeV is the nucleon mass. This mass selection provides a compromise value, being in closer agreement with the results of other microscopic calculations \cite[e.g.,][]{Baldo2014PhRvC}.

\begin{figure}[H]
\begin{adjustwidth}{-\extralength}{0cm}
\centering
\includegraphics[width=\fulllength]{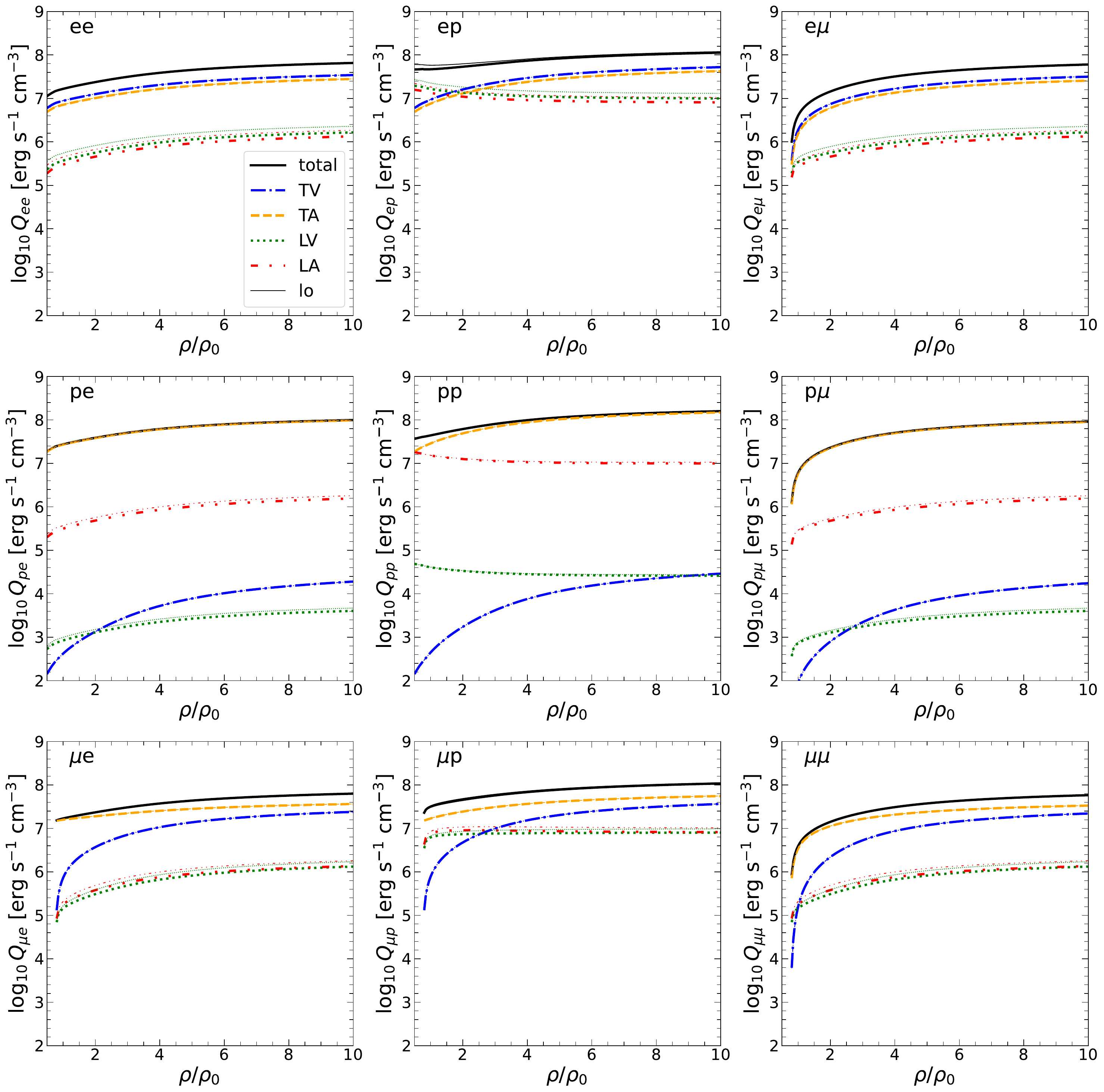}
\end{adjustwidth}
\caption{Partial contributions to bremsstrahlung emissivity due to electromagnetic collisions in normal NS cores with BSk 24 EoS as functions of density for $T=10^8$~K. Each panel marked $ij$ with $i,j=e,p,\mu$ corresponds to neutrino pair emission  by particle species $i$ scattered off particle $j$ as indicated in the panels. 
In each panel, different line types correspond to different contributions to emissivity as shown in legend and discussed in the text, while total contributions from each collision are shown with black solid lines. Thin lines of corresponding line types show the results of calculations in the lowest-order weak screening approximation.  
\label{fig:Qnorm_partial}}
\end{figure}

Let us start with normal (i.e. non-superconducting) matter. The total neutrino pair bremsstrahlung emissivity \eqref{eq:Qbr_sum} 
can be written as a sum of the partial contributions from each particle species pair given in Eq.~\eqref{eq:Q_12noscr}. In the table-shaped Fig.~\ref{fig:Qnorm_partial}, we plot the partial emissivities $Q_{ij}$, where $i,j=e,p,\mu$, as a function of normalized density $\rho/\rho_0$ for the temperature $T=10^8$~K. Each panel in Fig.~\ref{fig:Qnorm_partial} corresponds to a given pair of particles as indicated in the plot. Remind, that the matrix $Q_{ij}$ is not symmetric, since each term corresponds to neutrino pair being emitted by particle $i$, while particle $j$ plays the role of a passive scatterer. 
According to Eq.~\eqref{eq:Q_12noscr}, each partial emissivity contains four contributions LA, LV, TV, and TA from different combinations of transverse and longitudinal parts of the electromagnetic interaction with vector and axial-vector weak interaction channels. In each panel of Fig.~\ref{fig:Qnorm_partial},
these contributions are plotted with different line styles as indicated in the legend. With the solid lines in each panel we plot the total partial emissivities. Thick lines correspond to the results of exact numerical calculations according to Eqs.~\eqref{eq:FintLr}--\eqref{eq:FintTr} with complete expressions for the factors $F_{Lr}(v_{\mathrm{F}1},\kappa_1)$ and $F_{Tr}(v_{\mathrm{F}1},\kappa_1)$, while thin lines of respective types correspond to the lowest-order small-angle approximations, calculated via Eqs.~\eqref{eq:FintLr_lo}, \eqref{eq:IL2_lo}, \eqref{eq:FLV_lo}, and \eqref{eq:FLA_lo} for longitudinal contributions and via Eqs.~\eqref{eq:FTr_int_lo}, \eqref{eq:FTV_lo}, and \eqref{eq:FTA_lo} for transverse contributions, respectively. In almost all cases, the transverse contribution dominates the partial emissivity. This is due to considerably weaker screening in the transverse channel. The exceptions are given by processes of scattering off protons at low densities because the transverse channel in this regime is suppressed by a factor $v_{\mathrm{F}p}^2$ [cf. Eqs.~\eqref{eq:FTr_int_lo} and \eqref{eq:FintLr_lo}]. Emissivity through the (non-renormalized) vector and axial-vector currents is comparable, except for protons where the vector current contribution is suppressed by small values of the proton vector coupling constants; see Eq.~\eqref{eq:c_protons}.

Except for processes involving muons near the threshold of their appearance, all partial emissivities give comparable contributions to the total emissivity. We thus do not support the conclusion of Ref.~\cite{Kaminker1999} that emission from protons vanishes in the non-relativistic limit. 
It may seem from the $\mu e$ and $\mu p$ panels in Fig.~\ref{fig:Qnorm_partial} 
 that the emissivity by muons does not vanish at the threshold. This is because the drop in emissivity occurs when $k_{\mathrm{max}}$ in Eqs.~\eqref{eq:FintLr}--\eqref{eq:FintTr}, which is $2p_{\mathrm{F}\mu}$ in considered case, becomes smaller than the characteristic screening momentum. For the transverse channel, the characteristic screening scale $\Lambda_T$ is given by Eq.~\eqref{eq:LambdaTdef} and is very small, being proportional to $T^{1/3}$. Therefore, the interval of muon densities where its finite size plays a role is proportional to $T$ and as such is not resolved in Fig.~\ref{fig:Qnorm_partial} for the muon TA channel. The suppression for the vector current contribution in the transverse channel near the muon threshold is due to the additional factor $v_{\mathrm{F}1}^2$ in the non-relativistic limit for the factor $F_{TV}$, see Table~\ref{tab: Flims}.

Comparing thin and thick lines in Fig.~\ref{fig:Qnorm_partial}, we conclude that the lowest-order small angle approximation is perfect for the transverse channel, clearly due to very weak screening. It also has good performance in the longitudinal channel in the non-relativistic limit; see the proton row in Fig.~\ref{fig:Qnorm_partial}  and the low-density region of the muon row there. On the other hand, the lowest-order approximation overestimates the contribution in the longitudinal channel, especially in the ultrarelativistic limit. Mainly, this is due to the logarithmic term in the ultrarelativistic expression for the factors $F_{Lr}$, see Table~\ref{tab: Flims}. In contrast, in the non-relativistic limit, $F_{Lr}$ factors are just constants (do not depend on $\kappa_1$) and do not affect the small-angle approximation. As such, the only  difference between the lowest-order and exact calculations in  partial emissivities (thin and thick solid lines, respectively) visible at the resolution of Fig.~\ref{fig:Qnorm_partial}   is found for the $Q_{ep}$ term at low densities. 

\begin{figure}[H]
\includegraphics[width=0.7\textwidth
]{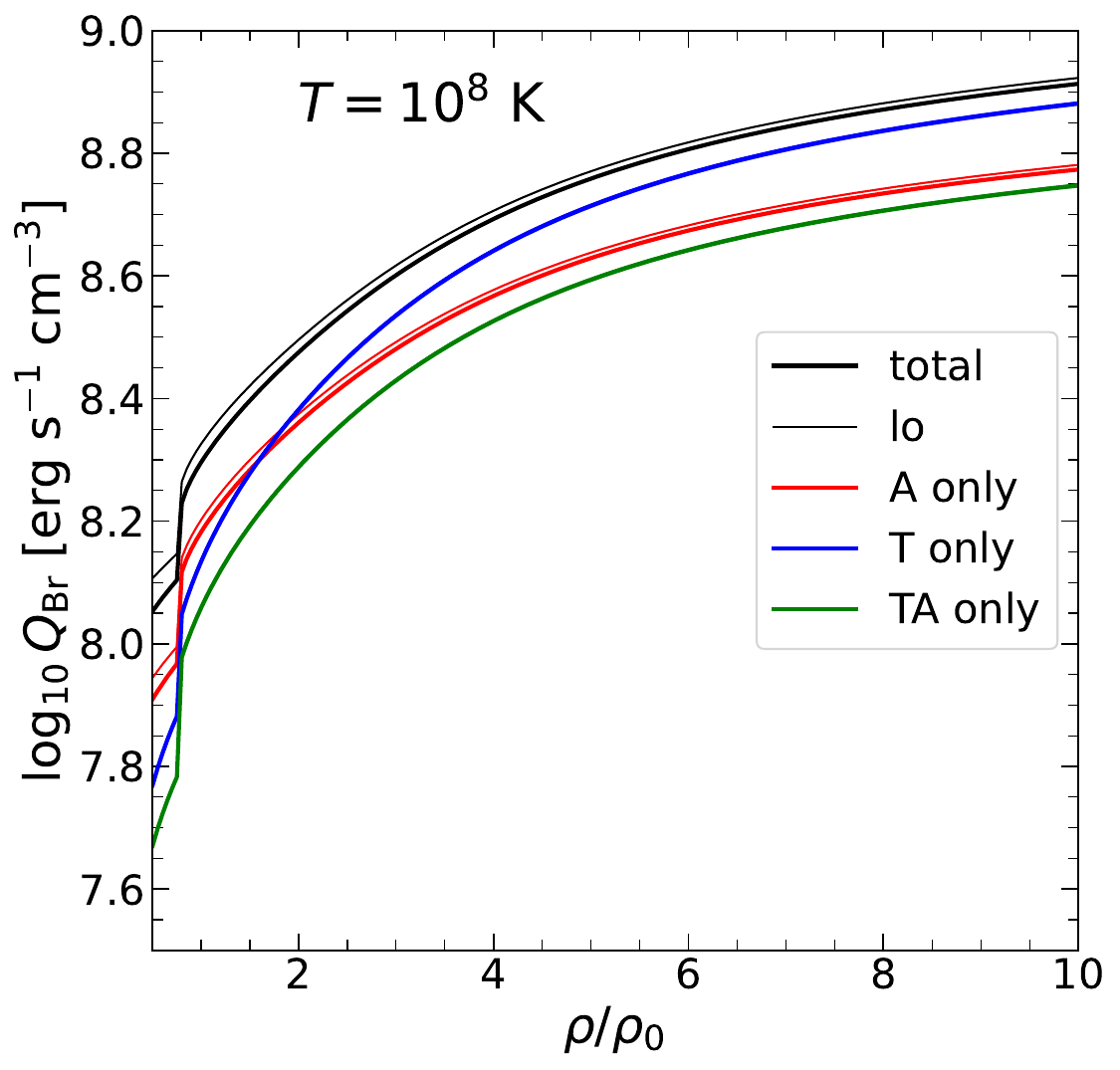}
\caption{Different approximations to the total bremsstrahlung emissivity due to electromagnetic interactions in non-superfluid BSk24 NS cores as a function of density for $T=10^8$~K. The results obtained via Eq.~\eqref{eq:Q_12noscr} with all contributions included are shown with the thick black line. The red line shows the contribution solely from the axial-vector channel. The blue  line corresponds to retaining only the transverse channel of the electromagnetic interaction, while the green line shows the contribution solely from the transverse electromagnetic channel to the axial-vector current emissivity. Thin lines of corresponding types represent the lowest-order weak screening approximation.  \label{fig:Qnorm_chan}}
\end{figure}   

In Fig.~\ref{fig:Qnorm_chan}, with a black solid line we show the total bremsstrahlung emissivity summed over all partial contributions as a function of density for $T=10^8$~K. Results of Refs.~\cite{Leinson1999,Leinson2001} suggest that neutrino emission through the vector current is suppressed. Following Ref.~\cite{Jaikumar2005PhRvD}, we set the vector coupling constants to zero and plot the corresponding results
in Fig.~\ref{fig:Qnorm_chan} with a thick red solid line. Suppression of the vector current contribution reduces the total emissivity by about 30\%. This will be our base result in what follows. In addition, in Fig.~\ref{fig:Qnorm_chan} we explore to what extent we can retain only the transverse contribution, as suggested in Ref.~\cite{Jaikumar2005PhRvD}. Corresponding results are shown with a green solid curve for axial-vector and transverse contribution and a blue solid curve for transverse vector and axial-vector contributions combined, respectively. We see that the longitudinal channel contributes up to 30\% at low densities (mainly due to the processes with protons), so it is advisable to include the longitudinal contribution despite the overall dominance of the transverse channel. Notice that due to the different temperature dependence in the longitudinal and transverse  contributions [cf. Eqs.~\eqref{eq:FintLr_lo} and \eqref{eq:FTr_int_lo}], the importance of the longitudinal contribution increases with temperature. A similar conclusion is reached when the shear viscosity of NS cores mediated by electromagnetic interactions is considered \cite{Shternin2022}, because expressions similar to Eqs.~\eqref{eq:FintLr}--\eqref{eq:FintTr} arise in that problem. 
Notice a prominent jump in the emissivity at the muon threshold. This is due to a large interaction range in the transverse channel of electromagnetic interaction. As discussed above, the size of the transition region from zero to the finite emissivity is very small.
Lowest-order small-angle approximations for all cases investigated in  Fig.~\ref{fig:Qnorm_chan} are shown with thin lines. Thin lines are not visible when the results contain only transverse contributions, and in the general case, the lowest-order result is also acceptable, especially provided the approximations used (i.e. neglect of the interference terms, etc.). We therefore will not discuss the difference between exact  and lowest-order results further.

\begin{figure}[H]
\includegraphics[width=0.7\textwidth]{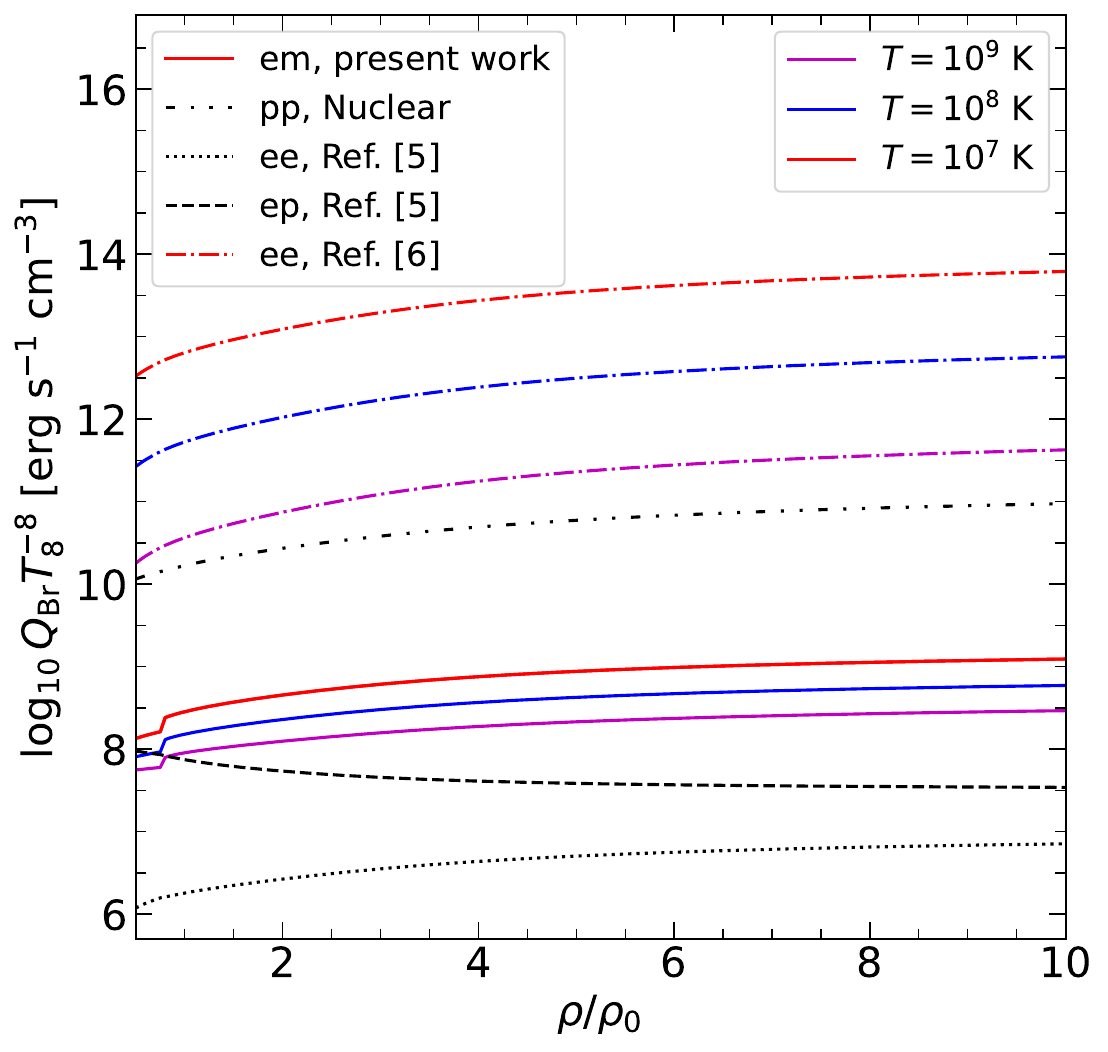}
\caption{Bermsstrahlung emissivity (scaled by $T_8^{-8}$) due to electromagnetic collisions as a function of density in normal NS cores with BSk24 EoS. Solid red, blue, and magenta lines shows the results of the present study for $T=10^7$~K, $10^8$~K, and $10^9$~K, respectively. Dash-dotted lines show the results of Ref.~\cite{Jaikumar2005PhRvD} for $ee$ bremsstrahlung for the same three values of temperature, and by dotted lines the results of Ref.~\cite{Kaminker1999} for $ee$ bremsstrahlung  are shown (in this case, combination  $Q^{ee}_{\mathrm{Br}}T_8^{-8}$ does not depend on temperature).  The black dashed line show the $ep$ bremmstrahlung emissivity according to Ref.~\cite{Kaminker1999}.}\label{fig:Qnorm_comp}
\end{figure}   
In Fig.~\ref{fig:Qnorm_comp}, we compare the results of our study with the results of previous works. We plot there the combination $Q_{\mathrm{Br}}T_{8}^{-8}$, where $T_8\equiv T/(10^8~\mathrm{K})$, which is temperature-independent for general bremsstrahlung processes \cite{YakovlevKaminker}, see also Eq.~\eqref{eq:Q_12noscr}. However, due to dynamical screening of the dominant transverse part of electromagnetic integration, this combination actually acquires a weak $T^{-1/3}$ temperature dependence. We plot $Q^{\mathrm{em}}_{\mathrm{Br}}T_{8}^{-8}$ for $T=10^7$, $10^8$, and $10^9$~K in Fig.~\ref{fig:Qnorm_comp} with red, blue, and magenta solid lines, respectively. For comparison, by dashed and dotted lines we plot the results of Ref.~\cite{Kaminker1999} for $ep$ and $ee$ bremsstrahlung, their eqs.~(40) and (51), respectively. In Ref.~\cite{Kaminker1999}, static Thomas-Fermi screening was employed in both longitudinal and transverse channels; therefore, first,  in their calculations, $Q^{ee}_{\mathrm{Br}}$ and $Q^{ep}_{\mathrm{Br}}$ are universally  proportional to $T^8$ and, second, the resulting emissivity is much smaller than those obtained in the present study.
The improvement of the result of Ref.~\cite{Kaminker1999} with correct plasma screening in the transverse channel was performed by Jaikumar et al. \cite{Jaikumar2005PhRvD}. We plot their results for $ee$ bremsstrahlung (eq.~(29) in Ref.~\cite{Jaikumar2005PhRvD}) with dash-dotted lines for the same three temperature values. In addition, with a double dot-dashed line we plot the neutrino  emissivity from the proton-proton bremsstrahlung via nuclear interactions calculated according to eq.~(167) in Ref.~\cite{YakovlevKaminker}. We further multiplied these results, which are based on calculations of Ref.~\cite{FrimanMaxwell1979ApJ}  and rely on the one-pion exchange description of nuclear interaction (with the addition of some short-range correlations)  by a factor of $0.25$ in order to mimic the effect of the use of the in-medium scattering matrix (see Ref.~\cite{VanDalen2003PhRvC} for details).  Proton-proton bremsstrahlung is apparently the weakest process mediated by nuclear interactions in normal matter \cite{YakovlevKaminker}. Other neutrino emission processes in nuclear sector, such as $np$ or $nn$ bremsstrahlung or modified and direct Urca processes, are even stronger and we do not show them in Fig.~\ref{fig:Qnorm_comp}.   When neutrons are not in the superfluid state, neutrino emission in electromagnetic processes is always much weaker than the emission rate of processes involving neutrons and can be safely neglected.

Jaikumar et al. \cite{Jaikumar2005PhRvD} obtained that neutrino emissivity in $ee$ collisions 
can be comparable to or exceed  neutrino emissivity in nucleon bremsstrahlung processes. Indeed, the dash-dotted lines in Fig.~\ref{fig:Qnorm_comp} overcome $pp$ bremsstrahlung emissivity by a couple  of orders of magnitude at low temperatures. {color{blue} This strong effect results from the  expression for the internal electron propagator employed in Ref.~\cite{Jaikumar2005PhRvD}, leading to the absence of the $k^2$ factor in the leading-order expressions for emissivity [cf. their Eq.~(15) with our Eqs.~\eqref{eq:FintLr}--\eqref{eq:FintTr}]. }
Our calculations result in a more modest increase of the emissivity with respect to the results of Ref.~\cite{Kaminker1999}.
In addition, the temperature dependence of emissivity is less prominent than found in Ref.~\cite{Jaikumar2005PhRvD}, since the dominant transverse contribution scales as $T^{-1/3}$ instead of $T^{-1}$ obtained by Jaikumar et al. \cite{Jaikumar2005PhRvD}.
Comparing solid lines in Fig.~\ref{fig:Qnorm_comp} with the double-dot-dashed line, we conclude that electromagnetic bremsstrahlung neutrino emission in NS cores can barely be important 
until all nucleons are in the well-developed paired state, so that all emission processes involving  nucleons (including Cooper pair formation emissivity) are suppressed.

We now turn to the situation where the all nucleonic species are in  paired states. The protons in NS cores are paired in the spin-singlet $^1S_0$ state \cite{LombardoSchulze2001LNP, SedrakianClark2019}, and in this case the temperature dependence of the energy gap can be approximated as \cite{Levenfish1994ARep} 
\begin{equation}\label{eq:gap}
\frac{\Delta}{T}=\sqrt{1-t}\left(1.456-\frac{0.157}{\sqrt{t}}+\frac{1.764}{t}\right),
\end{equation}
where $t=T/T_{Cp}$
To ensure that all processes involving protons are suppressed \cite{YakovlevKaminker}, we assume that $T/\Delta\lesssim 0.2$, or, in terms of critical temperature, $T\lesssim 0.35 T_{Cp}$.

\begin{figure}[H]
\begin{adjustwidth}{-\extralength}{0cm}
\centering
\includegraphics[width=\fulllength]{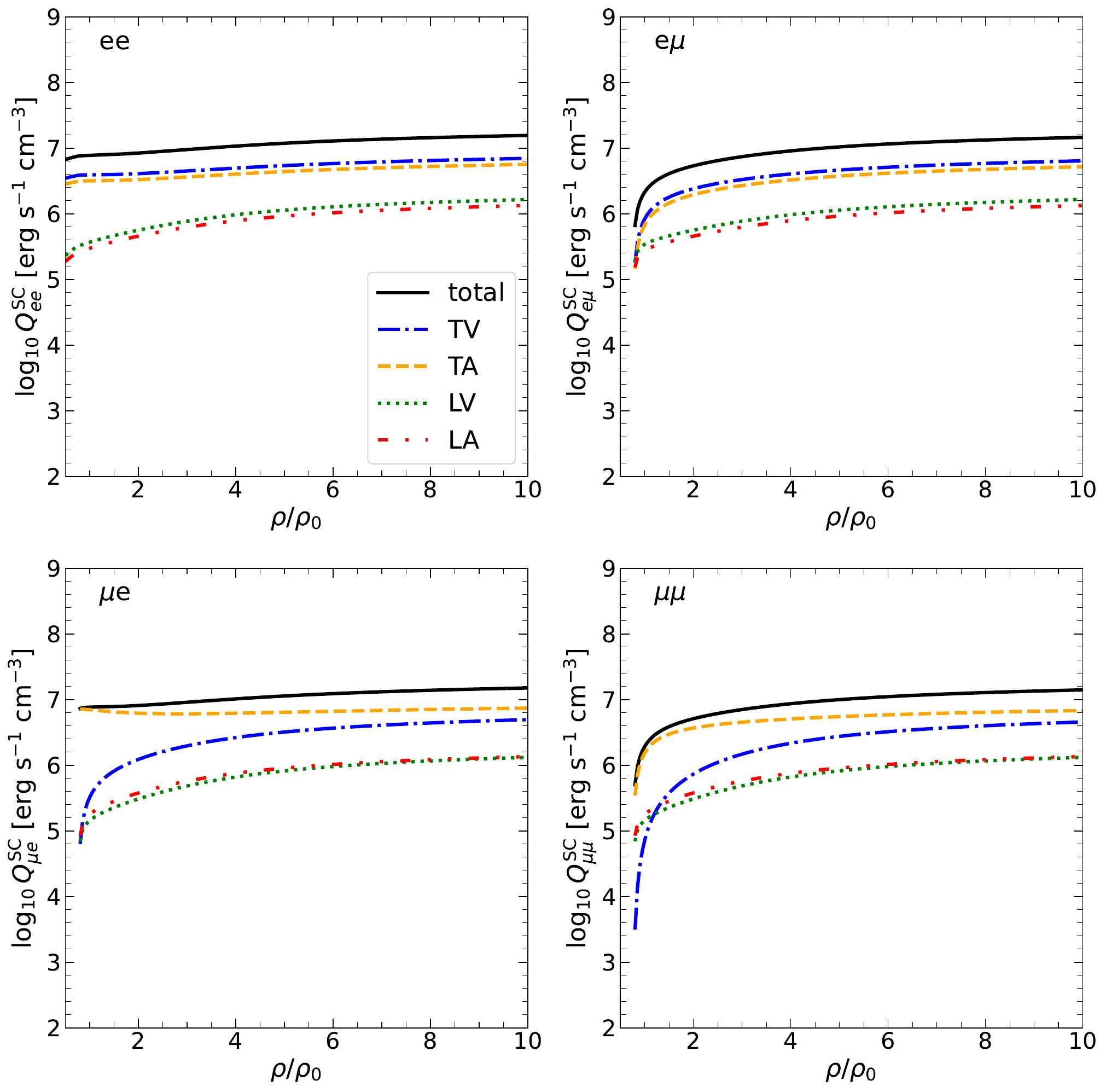}
\end{adjustwidth}
\caption{Partial contributions to bremsstrahlung emissivity due to electromagnetic collisions in superconducting NS cores with the BSk24 EoS, as functions of density for $T=10^8$~K and $T_{Cp}=10^{10}$~K. 
Each panel marked $ij$ with $i,j=e,\mu$ corresponds to neutrino pair emission  by particle species $i$ scattered off particle $j$ as indicated in the panels. 
In each panel, different line types correspond to different contributions to emissivity as shown in the legend and discussed in the text, while total contributions from each collision are shown with black solid lines (cf. Fig.~\ref{fig:Qnorm_comp}).  \label{fig:QSF_partial}}
\end{figure} 

In this case, the total neutrino pair bremsstrahlung emissivity is given by Eq.~(\ref{eq:Qbr_sum}), where the summation is carried over leptons -- electrons and muons. However, as discussed in Sec.~\ref{sec:screening_sc}, proton superconductivity modifies the character of plasma screening, therefore indirectly influencing  the lepton collision cross-sections and the transition rates. The partial contributions are still given by Eq.~\eqref{eq:Q_12noscr}, where the longitudinal contribution is not affected by proton pairing and is given by Eq.~\eqref{eq:FintLr} or Eq.~\eqref{eq:FintLr_lo} in the small-angle approximation. In contrast, the transverse contribution is modified, and one should employ Eq.~\eqref{eq:FintTr_SF_lo} in this case. Notice that since the characteristic transverse screening momentum $q_M$ in the superconducting matter is small, the lowest-order small-angle results still provide a good approximation. In Fig.~\ref{fig:QSF_partial}, which is similar to Fig.~\ref{fig:Qnorm_partial} for normal matter, we plot with solid lines the partial emissivities $Q_{ij}$, $i,j=e,\mu$ as functions of density $\rho/\rho_0$ for fixed value of temperature $T=10^8$~K and a  fixed density-independent proton critical temperature $T_{Cp}=10^{10}$~K. Contributions from different channels LV, LA, TV, and TA are shown with different line styles in each panel in Fig.~\ref{fig:QSF_partial} as indicated in the legend. One can notice that the transverse part of the interaction still dominates the rates; however, this dominance is not as strong as in normal matter. The reason is that the static screening scale in superconducting matter although being smaller than $q_{\mathrm{TF}}$, is considerably larger than the dynamical screening range in normal matter. 

\begin{figure}[H]
\includegraphics[width=0.7\textwidth]{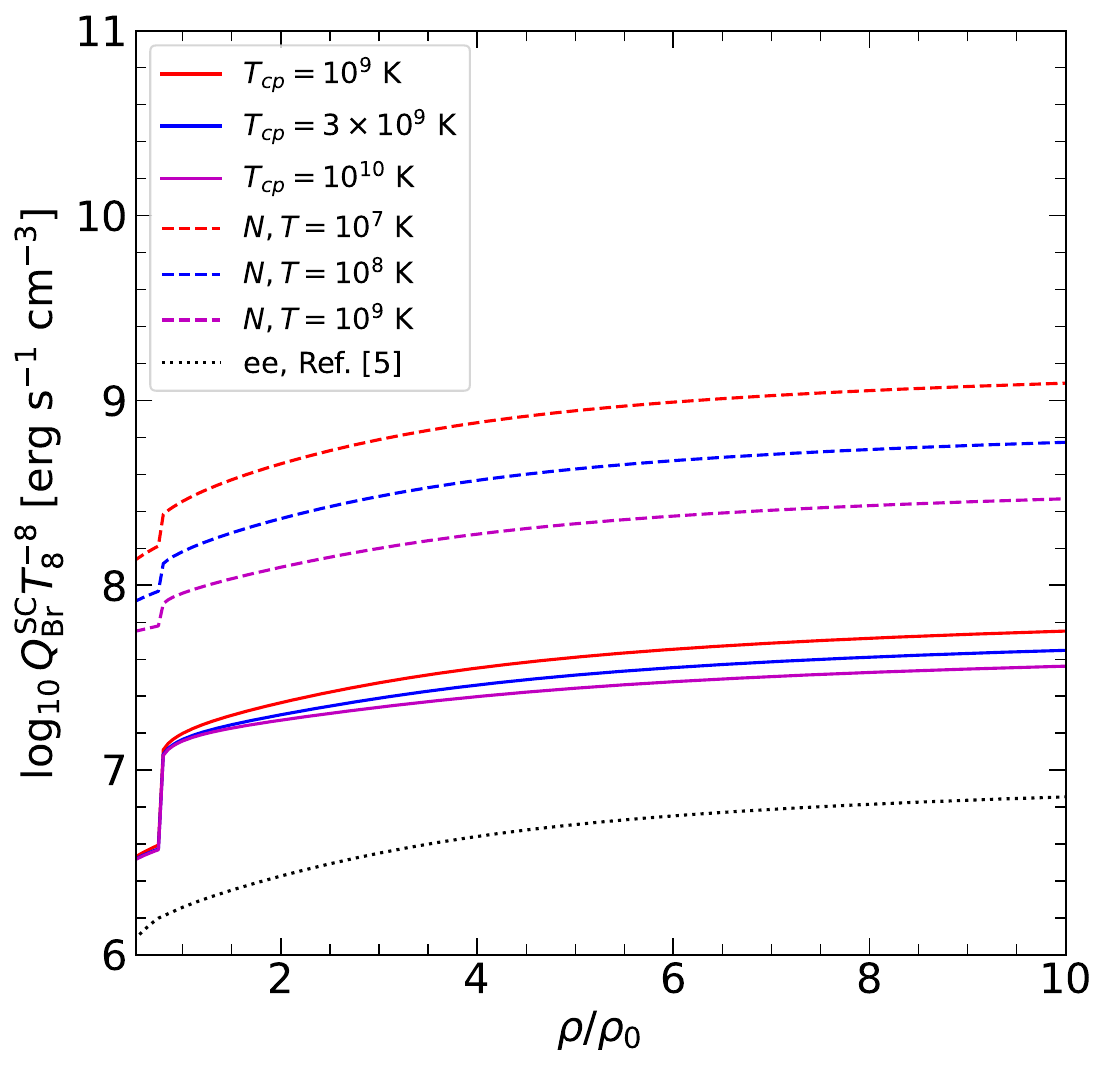}
\caption{Bermsstrahlung emissivity (scaled by $T_8^{-8}$) due to electromagnetic collisions as a function of density in superconducting NS cores with the BSk24 EoS. Red, blue, and magenta solid lines correspond to $T_{Cp}=10^9$~K, $3\times 10^9$~K, and $10^{10}$~K, respectively. Dashed red, blue, and magenta lines show emissivity in normal matter for  $T=10^7$~K, $10^8$~K, and $10^9$~K, respectively. For comparison, results of Ref.~\cite{Kaminker1999} for neutrino pair emissivity in electron-electron collisions is shown with the dotted line.\label{fig:QSF}}
\end{figure}   

Since the screening in superconducting matter is static, the standard temperature dependence $Q^{\mathrm{em}}_{\mathrm{Br}}\propto T^8$ is restored. However, the bremsstrahlung emissivity depends on the proton critical temperature $T_{Cp}$.
We illustrate this dependence in Fig.~\ref{fig:QSF}, where the total neutrino pair electromagnetic bremsstrahlung emissivity as a function of density $\rho/\rho_0$ is plotted with solid lines for $T=10^8$~K and three values of critical temperatures: $T_{Cp}=10^9$~K (red), $3\times 10^9$~K (blue), and $10^{10}$~K (magenta). As in normal matter, in this plot we retain only the axial-vector contribution, assuming that the vector current contribution is suppressed \cite{Leinson1999,Leinson2001}. For comparison, with dashed lines we plot the emissivities in normal matter for three values of temperature (as in Fig.~\ref{fig:Qnorm_comp}), and the results of Ref.~\cite{Kaminker1999} for electron-electron bremsstrahlung with a dotted line. The emissivity calculated in the present work is still larger than the results of Ref.~\cite{Kaminker1999}, which assume uniform Thomas-Fermi screening for both longitudinal and transverse channels, but the increase is not as large as observed for a normal matter (dashed lines). Nevertheless, electromagnetic bremsstrahlung in superconducting matter 
represents the residual emission which remains when the nucleon emission processes are suppressed and can be potentially important in practice. 
Figure~\ref{fig:QSF} shows that at low densities, the bremsstrahlung rate does not depend on the proton critical temperature. This is because at these densities the  London limit is applicable, where the screening is described by the Meissner mass $q_M$. At high densities, the Pippard limit is more applicable, and the bremsstrahlung emissivity starts to increase with decreasing critical temperature, roughly as $Q^{\mathrm{em}}_{\mathrm{Br}}\propto T_{Cp}^{-1/3}$. Notice that all these considerations are  valid when the superconductivity is well-developed, and this picture will be altered when the critical temperature becomes too low.

In the considered examples, the Dirac effective mass for protons was set to a constant value, $\widetilde{m}_p=0.8 m_N$. Using a different value, or a density-dependent effective mass $\widetilde{m}_p(\rho)$, would not qualitatively change the results. This parameter enters the derived expressions via the proton Fermi velocity $v_{\mathrm{F}p}$, Eq.~\eqref{eq:vF}, and directly affects the partial bremsstrahlung emissivities involving protons through the kinematic factors in Eqs.~\eqref{eq:M12nuphi_calc} and \eqref{eq:Q_12noscr}. These factors can change by at most a factor of a few between the non-relativistic and ultrarelativistic regimes. Therefore, variations in the proton effective mass would not allow the electromagnetic bremsstrahlung rate to become comparable to or exceed the nuclear proton-proton bremsstrahlung emissivity, provided the protons are unpaired.

Additionally, $v_{\mathrm{F}p}$, and hence $\widetilde{m}_p$, enters the expressions for the longitudinal polarization function \eqref{eq:Pil_norm} and the transverse polarization function in the superconducting case \eqref{eq:Pit_static}, affecting partial bremsstrahlung rates from all charged particle pairs. However, the effect of varying $\widetilde{m}_p$ is quite modest. In normal matter, the screening in the dominant transverse channel does not depend on $v_{\mathrm{F}p}$. The sub-dominant longitudinal contribution depends weakly on $v_{\mathrm{F}p}$ through the Thomas-Fermi screening momentum (Eq.~\eqref{eq:Pit_norm}); to leading order, this dependence is $\propto q_{\mathrm{TF}}^{-1}$ [see Eq.~\eqref{eq:FintLr_lo}]. In superconducting matter, the transverse screening does depend on $v_{\mathrm{F}p}$ [see Eqs.~\eqref{eq:Pit_static}--\eqref{eq:J_T0}]; however, this dependence vanishes to leading order in the Pippard limit. In the opposite, London limit, the transverse contribution to the bremsstrahlung rate is proportional to $v_{\mathrm{F}p}^{-1/2}$ to leading order [see Eqs.~\eqref{eq:FintTr_SF_lo} and \eqref{eq:Meissner}]. Similar conclusions have been reached for the related problem of lepton shear viscosity in superconducting NS cores \cite{Shternin2018}.

\section{Conclusions}\label{sec:concl}
We reconsidered the problem of neutrino pair bremsstrahlung emission accompanying charged particle collisions in nucleonic NS cores. We confirm the findings of Jaikumar et al.~\cite{Jaikumar2005PhRvD} that the main contribution to emissivity comes from the transverse channel of electromagnetic interaction. However, we obtained considerably lower values for $Q^{\mathrm{em}}_{\mathrm{Br}}$ than in Ref.~\cite{Jaikumar2005PhRvD} because the latter authors used a peculiar approximation for internal fermion propagator. With the  more appropriate form of the propagator, the differential cross-section in the transverse channel acquires additional small factor $k^2$ leading to a weaker $T^{-1/3}$ alteration of the temperature dependence  of bremsstrahlung emissivity than the $T^{-1}$ one found in Ref.~\cite{Jaikumar2005PhRvD}.
The practical expression for neutrino emissivity  of the considered process in nucleonic NS cores can be written as
\begin{adjustwidth}{-\extralength}{0cm}
\begin{eqnarray}
Q^{\mathrm{em}}_{\mathrm{Br}} &=& 2.1\times 10^{6}\ \left(\frac{n_B}{n_0}\right)^{2/3}T_8^8\sum_{ij=e,\mu,p} C_{iA}^2 Y_j^{2/3}\Bigg[\left(\frac{v_{\mathrm{F}j}}{c}\right)^{-2}
\left(\frac{1~\mathrm{fm}^{-1}}{q_{\mathrm{TF}}}\right)I_L^{(2)}\left(\frac{2\min \{p_{\mathrm{F}i},p_{\mathrm{F}j}\}}{\hbar q_{\mathrm{TF}}}\right)
F_{LA}^{lo}\left(\frac{v_{\mathrm{F}i}}{c}\right)\nonumber \\
&&+ 0.81 \left(\frac{1~\mathrm{fm}^{-1}}{\Lambda_T}\right)
F_{TA}^{lo}\left(\frac{v_{\mathrm{F}i}}{c}\right) \Bigg]~\mathrm{erg}~\mathrm{cm}^{-3}~\mathrm{s}^{-1},\label{eq:QNorm_practcal}
\end{eqnarray}
\end{adjustwidth}
where, following Refs.~\cite{Leinson1999, Leinson2001,Jaikumar2005PhRvD}, we retain only the axial-vector contribution, so that $C^2_{eA}=C^2_{\mu A}=0.75$, $C^2_{pA}=1.19$. In Eq.~\eqref{eq:QNorm_practcal}, $Y_j=n_j/n_B$ is  the fraction of particle species $j$, $n_B$ is the total baryon number density, and $n_0=0.16$~fm$^{-3}$ is the nuclear saturation number density. The factors $F_{LA}^{lo}$ and $F_{TA}^{lo}$ are given in Eqs.~\eqref{eq:FLA_lo}
 and \eqref{eq:FTA_lo}, respectively, and function $I_L^{(2)}(y_m)$ is given in Eq.~\eqref{eq:IL2_lo}. The characteristic screening momenta $q_{\mathrm{TF}}$ and $\Lambda_T$ can be calculated as
\begin{eqnarray}
    q_{\mathrm{TF}}&=&0.16\, \left(\frac{n_B}{n_0}\right)^{1/3}~\sqrt{\sum_{i=e,p,\mu} Y_i^{2/3} (v_{\mathrm{F}i}/c)^{-1}}\,  \mathrm{fm}^{-1},\label{eq:qTF_pract}\\
    \Lambda_T &=& 9.5\times 10^{-4}\,\left(\frac{n_B}{n_0}\right)^{1/3} T_8^{1/3}~\sqrt{\sum_{i=e,p,\mu} Y_i^{2/3}}\, \mathrm{fm}^{-1}.\label{eq:LambdaT_pract}
\end{eqnarray}
Fermi momenta in the same units are $p_{Fi}= 1.68 (n_B/n_0)^{1/3} Y_i^{1/3}$~fm$^{-1}$ and Fermi velocities can be calculated from Eq.~\eqref{eq:vF} or via $v_{\mathrm{F}i}=p_{\mathrm{F}i}/m_i^*$ if the Landau effective masses are available for the EoS used.

If either neutrons or protons are in normal states, the neutrino emission processes mediated by nuclear interaction dominate over Eq.~\eqref{eq:QNorm_practcal}. Only if $T\ll T_{Cp},\,T_{Cn}$ for both nucleonic species, neutrino emission in lepton collisions may become important. Strong proton superconductivity changes the dynamical character of transverse plasma screening from dynamic to static; still, the transverse contribution remains dominant in the emissivity.
Our analysis results in the following practical expression for bremsstrahnlung emissivity in superconducting NS cores
\begin{adjustwidth}{-\extralength}{0cm}
\begin{eqnarray}\label{eq:QSF_pract}
Q^{\mathrm{em}}_{\mathrm{Br}} &=& 2.1\times 10^{6}\ \left(\frac{n_B}{n_0}\right)^{2/3}T_8^8\sum_{ij=e,\mu} C_{iA}^2 Y_j^{2/3}\Bigg[\left(\frac{v_{\mathrm{F}j}}{c}\right)^{-2}
\left(\frac{1~\mathrm{fm}^{-1}}{q_{\mathrm{TF}}}\right)I_L^{(2)}\left(\frac{2\min \{p_{\mathrm{F}i},p_{\mathrm{F}k}\}}{q_{\mathrm{TF}}}\right)
F_{LA}^{lo}\left(\frac{v_{\mathrm{F}i}}{c}\right) \\
&&+ \left(\frac{1~\mathrm{fm}^{-1}}{q_M}\right) I_{TSF}^{(2)}\left(\frac{2\min \{p_{\mathrm{F}i},p_{\mathrm{F}j}\}}{\hbar q_{M}},A\right)
F_{TA}^{lo}\left(\frac{v_{\mathrm{F}i}}{c}\right) \Bigg]~\mathrm{erg}~\mathrm{cm}^{-3}~\mathrm{s}^{-1},
\end{eqnarray}
\end{adjustwidth}
where the summation is carried over the electron-muon subsystem, the longitudinal contribution is the same as in Eq.~\eqref{eq:QNorm_practcal}, while the function $I_{TSF}^{(2)}(y_m)$ is given by the fitting expression \eqref{eq:f2A}. The Messner mass $q_M$ and the parameter $A$ can be calculated as
\begin{equation}
        q_{M}=9.3\times 10^{-2}\, \left(\frac{n_B}{n_0}\right)^{1/3}~ Y_p^{1/3} \left(\frac{v_{\mathrm{F}p}}{c}\right)^{1/2}\,  \mathrm{fm}^{-1},
\end{equation}
\begin{equation}\label{eq:A_natural}
A=1.12 \left(\frac{Y_p}{0.1}\right)^{5/6}\left(\frac{n_B}{n_0}\right)^{5/6} \left(\frac{v_{\mathrm{F}p}}{c}\right)^{3/2} \frac{0.5~{\rm MeV}}{\Delta},
\end{equation}
where the value of the gap $\Delta$ is given in Eq.~\eqref{eq:gap}.
When the temperature is not so low, Eq.~\eqref{eq:QSF_pract} becomes invalid. In this case, however, other neutrino processes, such as $pp$ nuclear bremsstrahlung, are more important. Although $pp$ nuclear bremsstrahlung will be modified by proton pairing, at moderate temperatures this suppression   is not dramatic \cite{YakovlevKaminker}, so that this process will still be dominant. We can therefore recommend using Eq.~\eqref{eq:QNorm_practcal} in normal matter and switch to Eq.~\eqref{eq:QSF_pract} already at the transition temperature. When temperature decreases to such low values that the $pp$ bremsstrahlung is switched off, Eq.~\eqref{eq:QSF_pract} will provide an appropriate description of neutrino emission in electromagnetic collisions. At larger temperatures, $Q_{\mathrm{Br}}^{\mathrm{em}}$ is not important anyway, so the failure of Eq.~\eqref{eq:QSF_pract} is irrelevant. 

\vspace{6pt} 




\funding{
The work is funded by the baseline project FFUG-2024-0002 of the Ioffe Institute.}


\acknowledgments{The author acknowledges the contribution of A.S. Khlybov, who participated in the earlier stages of the work. The author thanks A.D. Kaminker and D.G. Yakovlev for discussions and D.D. Ofengeim for discussions and especially for the help with the construction of the fitting expression in Eq.~\eqref{eq:f2A}.}

\conflictsofinterest{The authors declare no conflicts of interest.} 


\abbreviations{Abbreviations}{
The following abbreviations are used in this manuscript:
\\

\noindent 
\begin{tabular}{@{}ll}
NS & Neutron Star\\
EoS & Equation of State\\
\end{tabular}
}

\appendixtitles{no} 
\appendixstart
\appendix
\section[\appendixname~\thesection]{}\label{app:MandF}
Performing explicit contractions in Eq.~\eqref{eq:M12}, one can express the result as a sum of contributions from different screening channels. In general we have longitudinal, transverse, and mixed term:
\begin{equation}
    \mathcal{M}_{12}=\mathcal{M}_L+\mathcal{M}_T+\mathcal{M}_{TL},
\end{equation}
where the longitudinal term is 
\begin{equation}
    \mathcal{M}_L = \frac{L_{L}}{|K^2-\Pi_L|^2}\approx \frac{L_{L}}{|{k}^2+q_{\mathrm{TF}}^2|^2},
\end{equation}
where
\begin{eqnarray}
    L_L &=& Q^2(4E_{\mathrm{F}1}^2-{k}^2)(4E_{\mathrm{F}2}^2-{k}^2)\left[\frac{{k}^2}{uv}(c_{1V}^2+c_{1A}^2)+m_1^2\left(\frac{1}{u}-\frac{1}{v}\right)^2(2c_{1A}^2-c_{1V}^2)\right]\nonumber\\
    &&-c_{1A}^2m_1^2(4E_{\mathrm{F}2}^2-{k}^2)\left[\frac{8Q^2{k}^2}{uv}+  \frac{(u-v)^2}{uv}-\frac{4 Q_0^2 {k}^2}{uv}\right].
\end{eqnarray}
The transverse term $\mathcal{M}_{T}$ can be written as
\begin{equation}
    \mathcal{M}_T=\frac{L_T}{|K^2-\Pi_T|^2}\approx\frac{L_T}{|{k}^2+\Pi_T|^2},
\end{equation}
where
\begin{eqnarray}
    L_T&=&Q^2\left((4p_{\mathrm{F}1}^2-{k}^2)(4p_{\mathrm{F}2}^2-{k}^2)\cos^2\phi +4{k}^2(p_{\mathrm{F}1}^2+p_{\mathrm{F}2}^2)\right)\nonumber\\
    &&\times\left[\frac{{k}^2}{uv}(c_{1V}^2+c_{1A}^2)+m_1^2\left(\frac{1}{u}-\frac{1}{v}\right)^2(2c_{1A}^2-c_{1V}^2)\right]\nonumber\\
    &&+c_{1A}^2m_1^2(4p_{\mathrm{F}2}^2+{k}^2)\left[\frac{8Q^2{k}^2}{uv}+  \frac{(u-v)^2}{uv}\right]\nonumber\\
    &&+\frac{4}{uv}{k}^2 m_1^2c_{1A}^2\left[(\bm{P}_{24}\bm{q})^2+{k}^2\bm{q}_{\perp}^2\right].  
\end{eqnarray}
Finally, the mixed term is
\begin{eqnarray}
    \mathcal{M}_{TL}&=&\Bigg\{-2Q^2E_1E_2\sqrt{(4p_{\mathrm{F}1}^2-{k}^2)(4p_{\mathrm{F}2}^2-{k}^2)}\cos\phi \nonumber \\
    &&\times\left[\frac{{k}^2}{uv}(c_{1V}^2+c_{1A}^2)+m_1^2\left(\frac{1}{u}-\frac{1}{v}\right)^2(2c_{1A}^2-c_{1V}^2)\right]\nonumber\\
    &&-\frac{16}{uv}{k}^2 m_1^2c_{1A}^2E_2Q_0(\bm{P}_{24}\bm{q})\Bigg\}\mathrm{Re}\frac{1}{(\bm{k}^2+\Pi_L)(\bm{k}^2+\Pi_T^*)}\nonumber\\
    &&+\frac{16Q^2}{uv}{k}^2 c_{1V}c_{1A}  E_2\big(\bm{k}[\bm{P}_{24}\times \bm{P}_{13}]\big) \mathrm{Im}\frac{1}{({k}^2+\Pi_L)({k}^2+\Pi_T^*)}.
\end{eqnarray}
Upon integration over $\phi$, the mixed term vanishes and one obtains
\begin{equation}
    \frac{1}{2\pi}\int_0^{2\pi} \mathrm{d}\phi \mathcal{M}=  \frac{L_{L}}{|\bm{k}^2+q_{\mathrm{TF}}^2|^2}+\frac{\widetilde{L}_{T}}{|\bm{k}^2+\Pi_{T}^2|^2},
\end{equation}
where
\begin{eqnarray}
    \widetilde{L}_T&=&\frac{1}{2}Q^2(4p_{\mathrm{F}1}^2+\bm{k}^2)(4p_{\mathrm{F}2}^2+\bm{k}^2)\nonumber\\
    &&\times\left[\frac{\bm{k}^2}{uv}(c_{1V}^2+c_{1A}^2)+m_1^2\left(\frac{1}{u}-\frac{1}{v}\right)^2(2c_{1A}^2-c_{1V}^2)\right]\nonumber\\
    &&+c_{1A}^2m_1^2(4p_{\mathrm{F}2}^2+\bm{k}^2)\left[\frac{8Q^2\bm{k}^2}{uv}+  \frac{(u-v)^2}{uv}+\frac{2\bm{k}^2}{uv}\bm{q}_\perp^2\right].
\end{eqnarray}

Performing the integrations over $\mathrm{d}\Omega_\nu$ and $\mathrm{d} x_\nu$, we arrive at Eq.~\eqref{eq:M12nuphi_calc}, where the complete expressions for the factors $F_{Lr}$ and $F_{Tr}$ are 
\begin{eqnarray}
     F_{LV} &=& (1-v_{\mathrm{F}1}^2\kappa_1^2)\Bigg[
     J_{uv}(v_{\mathrm{F}1},\kappa_1) +\frac{1-v_{\mathrm{F}1}^2}{2\kappa_1^2v_{\mathrm{F}1}^2}\left(J_{uv}(v_{\mathrm{F}1},\kappa_1)-J_{uv}(v_{\mathrm{F}1},0)\right)
     \Bigg],\\
    F_{TV}&=&\frac{1}{2}\frac{1+\kappa_1^2}{1-v_{\mathrm{F}1}^2\kappa_1^2} F_{LV},\\
    F_{LA}&=&(1-v_{\mathrm{F}1}^2\kappa_1^2)\Bigg[
     J_{uv}(v_{\mathrm{F}1},\kappa_1) -\frac{1-v_{\mathrm{F}1}^2}{\kappa_1^2v_{\mathrm{F}1}^2}\left(J_{uv}(v_{\mathrm{F}1},\kappa_1)-J_{uv}(v_{\mathrm{F}1},0)\right)
     \Bigg]\nonumber\\
     &&+ (1-v_{\mathrm{F}1}^2)\Bigg[
     -2J_{uv}(v_{\mathrm{F}1},\kappa_1)+\frac{1}{2}J_{uv}(v_{\mathrm{F}1},0)+J_\parallel(v_{\mathrm{F}1},\kappa_1) \Bigg],\\
    F_{TA}&=&\frac{1}{2}(1+\kappa_1^2)\Bigg[
     v_{\mathrm{F}1}^2J_{uv}(v_{\mathrm{F}1},\kappa_1) -\frac{1-v_{\mathrm{F}1}^2}{\kappa_1^2}\left(J_{uv}(v_{\mathrm{F}1},\kappa_1)-J_{uv}(v_{\mathrm{F}1},0)\right)
     \Bigg]\nonumber\\
      &&+ (1-v_{\mathrm{F}1}^2)\Bigg[
      2J_{uv}(v_{\mathrm{F}1},\kappa_1)-\frac{1}{2}J_{uv}(v_{\mathrm{F}1},0)+\frac{1}{2}J_{\perp}(v_{\mathrm{F}1},\kappa_1) \Bigg].
\end{eqnarray}
In these expressions, the following integrals are introduced
\begin{eqnarray}
   J_{uv}(v_{\mathrm{F}1},\kappa_1) &=& 4E_1^2Q_0^2 \int\frac{\mathrm{d}\Omega_\nu}{4\pi}\int_0^1\frac{x_\nu^2 (1-x_\nu^2)\mathrm{d}x_\nu}{uv}\nonumber\\
   &=&\frac{2(1-\beta^2)^{3/2}}{3\alpha\beta^4} \mathrm{ArcTanh}\frac{\alpha}{\sqrt{1-\beta^2}} \nonumber\\
     &&-\frac{1}{3\beta^4v_{\mathrm{F}1}^3} \Big(-v
     _{\mathrm{F}1}\beta^2 +\left(\beta^2+v_{\mathrm{F}1}^2 (2-3\beta^2)\right)\mathrm{ArcTanh}\, v_{\mathrm{F}1}\Big),\\
     J_{uv}(v_{\mathrm{F}1},0) &=& \frac{1}{v_{\mathrm{F}1}^4}-\frac{2}{3v_{\mathrm{F}1}^2}-\frac{(1-v_{\mathrm{F}1}^2)}{v_{\mathrm{F}1}^5}\mathrm{ArcTanh}\, v_{\mathrm{F}1},\\
     J_\parallel&=&4E_1^2Q_0^2 \int\frac{\mathrm{d}\Omega_\nu}{4\pi}\int_0^1\frac{x_\nu^2 \mathrm{d}x_\nu}{uv}\nonumber\\
     &=&-\frac{(1-\beta^2)^{1/2}}{\alpha\beta^2} \mathrm{ArcTanh}\frac{\alpha}{\sqrt{1-\beta^2}} +\frac{1}{\beta^2v_{\mathrm{F}1}} \mathrm{ArcTanh}\, v_{\mathrm{F}1},\\
     J_\perp&=&4E_{1}^2Q_0^2 \int\frac{\sin^2\theta_\nu \mathrm{d}\Omega_\nu}{4\pi}\int_0^1\frac{x_\nu^4 \mathrm{d}x_\nu}{uv}\nonumber\\
     &=&\frac{1}{3v_{\mathrm{F}1}^2}-\frac{1}{2v_{\mathrm{F}1}^4}+\frac{(1-v_{\mathrm{F}1}^2)\mathrm{ArcTanh}\, v_{\mathrm{F}1}}{2v_{\mathrm{F}1}^5}\nonumber\\
     &&-\frac{\sqrt{1-\beta^2}(2+\beta^2)}{3\alpha\beta^4}\mathrm{ArcTanh}\frac{\alpha}{\sqrt{1-\beta^2 }}\nonumber\\
    &&+\frac{1}{3\beta^4 v_{\mathrm{F}1}^3}\left[-v_{\mathrm{F}1}\beta^2+(2v_{\mathrm{F}1}^2+\beta^2)\mathrm{ArcTanh}\,v_{\mathrm{F}1}\right],
\end{eqnarray}
where $ \alpha=v_{\mathrm{F}1} \kappa_1$ and $\beta=v_{\mathrm{F}1}\sqrt{1-\kappa_1^2}$.

\section[\appendixname~\thesection]{}\label{app:num_trans}
Let us give details on numerical integration in Eq.~\eqref{eq:FintTr}.
It is convenient to express $\mathcal{F}_{Tr}$ as 
\begin{equation}
    \mathcal{F}_{Tr}=\frac{4}{\Lambda_T}\zeta_{-1/3}\left(p_{\mathrm{F}2}^2 I_{Tr12}^{(2)}+p_{\mathrm{F}1}^2 I_{Tr12}^{(4)}\right),
\end{equation}
where 
\begin{eqnarray}
    I_{Tr12}^{(n)}&=&\frac{y_T^{n-4}}{\zeta_{-1/3}}\int\limits_{-\infty}^{+\infty} \mathrm{d}\varpi \int\limits_0^\infty \mathrm{d}w\, 
    \mathcal{Z}(w,\varpi)
    \int_0^{y_{Tm}} \frac{y^{n-2}\,\mathrm{d}y}{|y^2+i\varpi/y|^2}F_{Tr}(v_{\mathrm{F}1},y y_{T1}),
\end{eqnarray}
$y_T=\Lambda_T/(2p_{\mathrm{F}1})$, and $y_{Tm}=k_{\mathrm{max}}/\Lambda_T$.
The energy integration may be performed by introducing the function
\begin{equation}
    K(\Upsilon)\equiv\int\limits_{-\infty}^{+\infty} \mathrm{d}\varpi \int\limits_0^\infty \mathrm{d}w\,
    \frac{\mathcal{Z}(w,\varpi)}{\varpi^2+\Upsilon}.
\end{equation}
Asymptotically, $ K(\Upsilon)\to \zeta_0/\Upsilon$, $\Upsilon\to +\infty$ and $ K(\Upsilon)\to 8\pi^7 /(63\sqrt{\Upsilon})$, $\Upsilon\to 0$.
The simple fitting expression 
\begin{equation}
    K(\Upsilon)=\frac{8\pi^7}{63\sqrt{\Upsilon}}\frac{1+0.00813\sqrt{\Upsilon}}{1+0.2132\sqrt{\Upsilon}+0.0208\Upsilon}
\end{equation}
works well allowing to express the transverse integrals as 
\begin{equation}
    I_{Tr12}^{(n)}=\frac{y_T^{n-4}}{\zeta_{-1/3}}\int_0^{y_{Tm}}\mathrm{d}y\, y^{n} K(y^6) F_{Tr}(v_{\mathrm{F}1},y y_{T1})
\end{equation}
and calculate them numerically.

\begin{adjustwidth}{-\extralength}{0cm}

\reftitle{References}




\PublishersNote{}
\end{adjustwidth}
\end{document}